\documentclass[aps,onecolumn,a4paper]{revtex4}
\usepackage[colorlinks=true, pdfstartview=FitV, linkcolor=blue, citecolor=red, urlcolor=magenta, breaklinks=true]{hyperref}
\usepackage{graphicx}
\usepackage[all]{xy}
\usepackage{amsmath}
\usepackage{amssymb}
\usepackage{color}
\usepackage{subeqnarray}
\usepackage{subfigure}
\newcommand{\bb}{\bibitem}
\newcommand{\bes}{\begin{subequations}}
\newcommand{\ees}{\end{subequations}}
\def\ben{\begin{eqnarray}}
\def\een{\end{eqnarray}}
\newcommand{\bens}{\begin{subeqnarray}}
\newcommand{\eens}{\end{subeqnarray}}
\def\be{\begin{equation}}
\def\ee{\end{equation}}
\def\tanh{\text{tanh}}
\def\arctanh{\text{arctanh}}
\def\cos{\text{cos}}
\def\arccos{\text{arccos}}
\def\sin{\text{sin}}
\def\sech{\text{sech}}
\def\cotg{\text{cotg}}
\def\cosh{\text{cosh}}
\def\sinh{\text{sinh}}
\def\arcsinh{\text{arcsinh}}
\def\sn{\text{sn}}
\def\sd{\text{sd}}
\def\cn{\text{cn}}
\def\dn{\text{dn}}

\def\e{\text{e}}
\def\o{\text{o}}

\begin{document}

\title{Kinks and branes in models with hyperbolic interactions} 
\author{D. Bazeia\footnote{Corresponding author. Email: bazeia@fisica.ufpb.br}} 
\author{Elisama E. M. Lima}
\author{L. Losano}  
\affiliation{Departamento de F\'\i sica Universidade Federal da Para\'iba, 58051-900 Jo\~ao Pessoa PB, Brazil}
\date{\today}
\begin{abstract}
In this work we investigate several models described by a single real scalar field with non-polynomial interactions, constructed to support topological solutions. We do this using the deformation procedure to introduce a function which allows to construct two distinct families of hyperbolic potentials, controlled by three distinct parameters, in the standard formalism. In this way, the procedure allows us to get analytical solutions, and then investigate the energy density, linear stability and zero mode. We move on and introduce a non-standard formalism to obtain compact solutions, analytically. We also investigate these hyperbolic models in the braneworld context, considering both the standard and non-standard possibilities. The results show how to construct distinct braneworld models which are implemented via the first order formalism and are stable against fluctuation of the metric tensor.
\end{abstract}

\maketitle

\section{Introduction}
\label{sec:intro}

Topological defects have been investigated for many years due to their importance in high energy physics \cite{r1,r11,r12,DM1,DM2,DM3} and in condensed matter physics \cite{CMab1,CMab2,CMab3,CMa1,CMa2,CMa3,CMa4,CMa5,CMa6}, and in other branches of non-linear science \cite{NLC1,NLC2}. Particularly, they appear during phase transitions in the evolution of early universe or between interfaces separating distinct regions in space \cite{r1,r11,r12} and, in this context, there have been suggested that some characteristics of the universe today may be intimately tied to such topological structures. Another interesting issue is addressed to condensed matter systems where, for example, in Ref.~\cite{Parkin190} one enquired the development of magnetic memory at the nanometric scale through the study of formation and propagation of domain walls in magnetic nanowires.

The most common topological structures are one-dimensional static solutions of the equations of motion; they are called kinks and appear in models described by scalar fields exhibiting spontaneous symmetry breaking. Because of its simplicity, a kink can be used to provide information on the behavior of physical systems in several important subject of physics \cite{Basar,Izquierdo,Romanczukiewicz,Pyka,brane1,brane2,braneB1,braneB2,braneB3}. For instance, kink structures may be used to model exact solutions for the one dimensional Bogoliubov-de Gennes and the Eilenberger equations of superconductivity \cite{Basar}, to schematize a moduli space of non-relativistic solitary waves at the long wavelength limit of ferromagnetic spin chains \cite{Izquierdo}, in the study of kink-antikink pairs production over collision of particle-like states \cite{Romanczukiewicz}, and more recently in Ref.~\cite{Pyka} is demonstrated  the formation of stable kinks during a structural phase transition in ion Coulomb crystals. Additionally, kinks may be used in braneworld models with a single extra dimension of infinite extent, as investigated in \cite{brane1,brane2,braneB1,braneB2,braneB3}, for instance.

In order to further contribute to the knowledge based on topological defects in field theories, we develop in this paper new families of hyperbolic models for real scalar fields not yet studied in the literature. Starting from the well-known $\phi^4$ theory, we benefit from the deformation procedure developed in Ref.~\cite{AA1,AA2,AA3} to find analytical topological solutions for the new models, and through basic manipulations we calculate some of their properties, such as stability and zero modes. This part of the investigation follows the lines of the work \cite{PP}, where one developed families of polynomials potentials and their corresponding defect structures; however, here we concentrate on scalar fields described by non-polynomial self-interactions, or more specifically by hyperbolic interactions with higher and higher powers. In general, it is a laborious task to find analytical solutions for such systems, but in this work we show how to obtain results concerning the presence of exact solutions, in a way that follows by direct use of the deformation procedure introduced in \cite{AA1,AA2,AA3}. 

Systems described by  non-polynomial interactions are widely studied in the literature, among them stand out the sine-Gordon models \cite{Sine1,Sine2,Garcia,Roldao,Alonso,Cruz1,Cruz2}. Such models are employed to study defect structures \cite{Sine1,Sine2,Garcia}, in investigations involving braneworlds in flat and curved spacetime \cite{Roldao}, entanglement in chain of particles \cite{Alonso}, localization of gravity and gauge fields \cite{Cruz1,Cruz2}. In this regard, the models studied in this work open the way to new investigations, and here we recall that the sinh-Gordon model is useful in a diversity of contexts, in particular to provide solutions of minimal surfaces in an anti-de Sitter space \cite{MS1a,MS1b,MS2} and to investigate integrability \cite{IN1,IN2,IN3}. Related to issues of direct interest to string theory, the current study may also be used within the AdS/CFT correspondence scenario, to investigate problems within the holographic cosmology environment \cite{HC1,HC2}. We also believe that the new models and their corresponding kinklike solutions will stimulate new researches, in particular on kink-antikink and multi-kink collisions, to see if new effects appear due to the hyperbolic interactions. This route to investigate collisions may be implemented following the recent studies \cite{C0a,C0b,C1a,C1b,C2,C3,C4}.

We go on and consider hyperbolic potentials in a non-standard perspective, starting from a modification on the kinematics of the system. Usually, such modifications introduce new kinds of non-linearities whose study is very complicate; however, here the equations of motion were solved exactly and it was possible to verify the presence of kinklike defects with compact support, known in the literature as compactons  \cite{Rosenau}. These structures are located exactly on some compact region of space and may appear in field theories with modified kinematics including non-linear dispersion  \cite{Babichev1,Babichev2,adam1,adam2,FOFE,FOGD,BGD,DMGD}, or still they can be manifested in standard models under specific conditions \cite{Arodz1,Arodz2,cstan1,cstan2}. By the way, compact structures have been studied in several distinct ways in the recent works \cite{BV,Ambroise, CE,Lima1, Lima2}.

Moreover, we also investigate the possibility to incorporate the explored models into a warped geometry with a single extra dimension of infinite extent, in distinct braneworld scenarios \cite{RS,Fre}. In this way, we cover situations related to scalar fields with usual and modified dynamics, including the presence of first order equations to simplify the investigation, inducing the presence of analytical solutions. To do this, we organize this work as follows. In the next section, we briefly review the standard formalism of a single real scalar field in $(1,1)$ spacetime dimensions. In Sec.~\ref{sec-2} we obtain new families of hyperbolic potentials, using the deformation method, as well as their properties such as kink solutions, energy density, stability potential and zero mode. In Sec.~\ref{sec-3} we deal with generalized models, modifying the kinematic term of the Lagrange density to study the possibility of the existence of compact structures. Through the study of a new deformation developed in the case of non-standard kinematics, we were able to get new compact solutions. Furthermore, in Sec.~\ref{sec-4} we consider the standard and non-standard models in the five-dimensional braneworld scenario with an infinite extra dimension, showing how the braneworld models behave for results obtained under the first order framework. We end the work including our comments and conclusions in Sec.~\ref{sec-com}.

\section{Standard Formalism}\label{sec-1}

We start our investigation from the general Lagrange density
describing a relativistic system driven by a single real scalar field
\be
\label{lagran}
{\mathcal L}=\frac{1}{2}\partial_\mu \phi \partial^\mu \phi-V(\phi),
\ee
where $V(\phi)$ is a potential which specifies self-interactions of the scalar field $\phi$. Our notation is usual in natural units $\hbar=c=1$. If we assume the two-dimensional spacetime with the metric $(+,-)$,  the field is dimensionless and the spacetime coordinates have dimension inverse of energy. Here, however, we shall work with dimensionless field and spacetime coordinates. The  equation of motion is given as follow
\ben
\label{eom}
\frac{\partial^2 \phi}{\partial t^2}- \frac{\partial^2 \phi}{\partial x^2}+\frac{d V}{d\phi}&=&0.
\een
From the point of view of topological defects, it is interesting to assume that $V(\phi)$ 
supports at least two neighbouring minima which characterize a topological sector. For static configurations, $\phi(x)$, a topological solution connects these minima and solve the equation
\be 
\frac{d^2\phi}{dx^2}=\frac{dV}{d\phi},
\ee
which can be written as
\be 
\frac12\left(\frac{d\phi}{dx}\right)^2=V(\phi),
\ee
since the solution has to have finite energy. In this case, the energy density is given by
\be
\rho(x)=\frac{1}{2}\left(\frac{d\phi}{dx}\right)^2+V(\phi)=\left(\frac{d\phi}{dx}\right)^2=2V(\phi).
\ee

To evaluate quantum effects or the linear stability, the static field $\phi(x)$ is submitted to small fluctuations of the type 
$\phi(x,t)= \phi(x) +\eta(x,t)$. Substituting this into (\ref{eom}) and expanding up to first-order in $\eta(x,t)$, and so using $\eta(x,t)=\sum_n\eta_{n}(x)\cos(\omega_{n} t)$, we get a Schr$\ddot{\o}$dinger-like equation $H\eta_{n}(x)=\omega_{n}^{2}\eta_{n}(x)$, such that
\be
\label{neq}
\left(-\frac{d^2}{dx^2}+U(x)\right)\eta_{n}(x)=\omega_{n}^{2}\eta_{n}(x),
\ee
where 
\be
\label{pet}
U(x)=\left.\frac{d^2V}{d\phi^2}\right|_{\phi(x)}.
\ee
It is possible to verify that the Eq.~\eqref{neq} provides at least one bound state, that is the bosonic zero mode given by
\be\label{zm}
\eta_0(x)=\frac{d\phi}{dx}.
\ee
For the situation in which the potential is written as $V(\phi)=(1/2)W_{\phi}^{2}$, where $W=W(\phi)$ and $W_{\phi}=dW/d\phi$, the equation of motion for the static field can be reduced to first-order
\be
\label{foe}
\frac{d\phi}{dx}=W_{\phi}.
\ee
So the Hamiltonian  \eqref{neq} can be written as $H=A^\dag A$, where
\be
A^\dag=-\frac{d}{dx} - W_{\phi\phi}\,\,\,\,\, \mbox{and} \,\,\,\,\,
A=\frac{d}{dx} -W_{\phi\phi}. \nonumber
\ee
This shows that the eigenvalues of $H$ are positive defined. Thus, the solution of the first-order equation \eqref{foe} is linearly stable. 

To create new models for real scalar fields we can use the deformation procedure \cite{AA1,AA2,AA3}. This method consists of finding a deforming function, $f(\phi)$, which connects a potential $U(\chi)$, for which one knows the solutions, to another one, for which one wishes to find the solutions, $V(\phi)$. In the standard scenario, these two potentials are related by 
\be
V(\phi)=\frac{U(\chi\rightarrow f(\phi))}{(df/d\phi)^2}.
\ee
This procedure allows that we get solutions for new potential using the inverse of the deformation function, $\phi(x)=f^{-1}(\chi(x))$, with $\chi(x)$ being the solution of the starting model described by $U(\chi)$. We exemplify this explicitly in the next section.

\section{New families of models} \label{sec-2}

For our propose, we start with the well known $\chi^4$ model, and we search
for new  hyperbolic models which support topological solutions. The potential can be written as
\be
\label{chi4}
U(\chi)=\frac{1}{2}(1-\chi^2)^2.
\ee
This model has two minima $\bar{\chi}_\pm=\pm 1$ and static topological solutions $\chi(x)=\pm \tanh(x)$. 

\subsection{First family of models}\label{family-1}

We now propose a deformation function expressed as
\be\label{def1}
f^{(1)}(\phi)=\cos\left[a \,\arccos\left(\sqrt{n}\,\tanh(\phi)\right)-m\pi\right]\,,
\ee
where $a$ is a non-null positive integer constant, $m$ is positive integer, and $n$ is a real parameter bigger than one. The deformed potential can be written in the form
\be
V_a^{(1)}(\phi)=\frac{1}{2na^2}\frac{\left(1-(n-1)\sinh^2(\phi)\right)}{\sech^2(\phi)}\left(1-[f^{(1)}(\phi)]^2\right)\,.
\ee
This potential can be written in terms of Chebyshev polynomials
\be
\label{pott}
V_a^{(1)}(\phi)=\frac{\left(1-(n-1)\sinh^2(\phi)\right)^2}{2na^2}U_{a-1}^{2}\left[\sqrt{n}\,\tanh(\phi)\right]. \\
\ee
The Chebyshev polynomials have the general form  
\be
\label{Chebyshev}
U_{b}\{\sigma\}=\frac{\sin\{(b+1)\, \arccos(\sigma)\}}{\sin\{\arccos(\sigma)\}},
\ee
where $b$ takes non-negative integer values and the argument $\sigma$ is unrestricted.

\begin{figure}
\includegraphics[width=7.5cm,height=6cm]{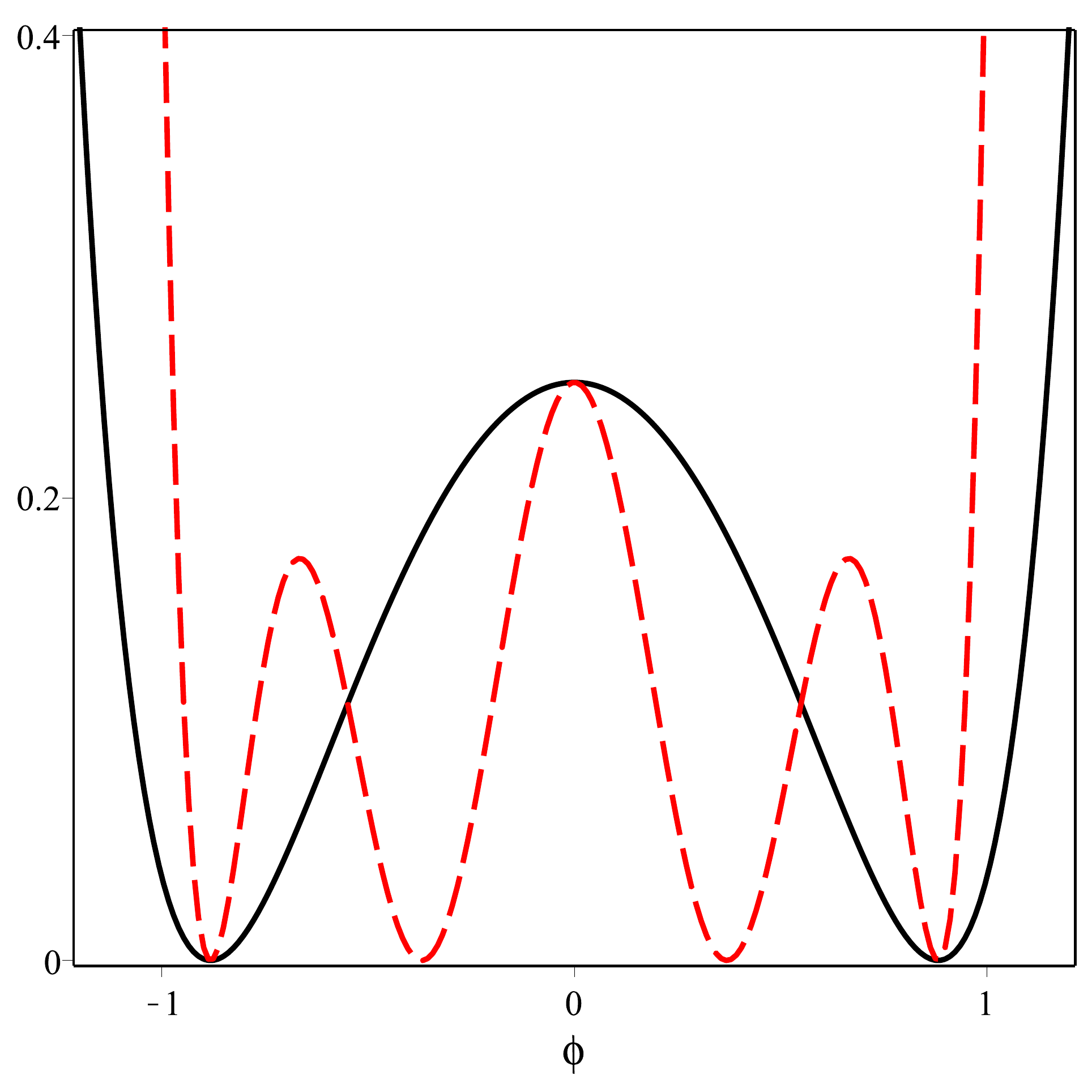}
\includegraphics[width=7.5cm,height=6cm]{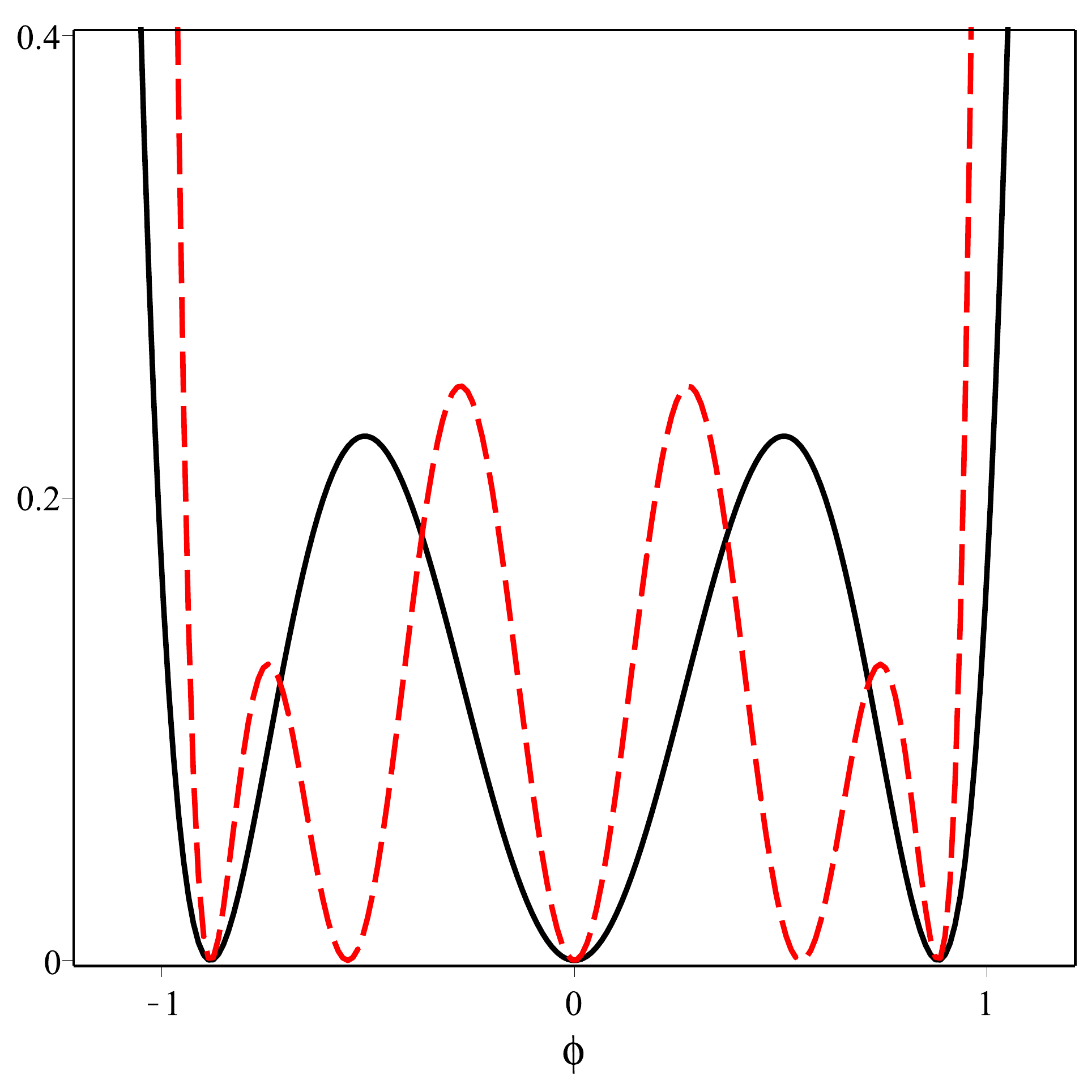}
\caption{The potentials \eqref{pott} displayed as $a^2V_a^{(1)}(\phi)$ for $n=2$. In the left panel one shows $V_1^{(1)}(\phi)$ and $V_3^{(1)}(\phi)$, depicted with solid (black) and dashed (red) lines, respectively. In the right panel one depicts $V_2^{(1)}(\phi)$ and $V_4^{(1)}(\phi)$ with solid (black) and dashed (red) lines, respectively.}
\label{fig:1}
\end{figure}

Explicit results of $V_a^{(1)}(\phi)$  allow us to visualize clearly the shape of the new family of hyperbolic potentials. For instance, when $a= 1, 2, 3, 4$ we have
\ben\label{vf1}
V_1^{(1)}(\phi)&=&\frac{1}{2n}\left(1-(n-1)\,\sinh^2(\phi)\right)^2;  \\
V_2^{(1)}(\phi)&=&\frac{1}{2}\tanh^2(\phi)\left(1-(n-1)\,\sinh^2(\phi)\right)^2; \\
V_3^{(1)}(\phi)&=&\frac{1}{18n}\sech^4(\phi)\left(1-(4n-1)\,\sinh^2(\phi)\right)^2\left(1-(n-1)\,\sinh^2(\phi)\right)^2; \\
V_4^{(1)}(\phi)&=&\frac{1}{2}\tanh^2(\phi)\sech^4(\phi)\left(1-(2n-1)\,\sinh^2(\phi)\right)^2\left(1-(n-1)\,\sinh^2(\phi)\right)^2. \:\:\:\:\:\:\:\:\:\:\:\:\:\:\:\:
\een
We illustrate these potentials in the Fig. \ref{fig:1}, which shows how the increase of $a$ leads to the appearance of new topological sectors.
In fact, the minima of the potentials are given by 
\be
\bar{\phi}=\arctanh\left[\frac{1}{\sqrt{n}}\cos\left(\frac{k\pi}{a}\right)\right]
\ee
where $k=0,..,a$. Each value of $a$ leads to $a+1$ minima. 

Assuming static configurations, the equation of motion is given by 
\ben
\frac{d^2\phi}{d x^2}&=&\left\{ \frac{1}{a^2n}\tanh(\phi)\left[(2-n)-2(n-1)\sinh^2(\phi)\right]U_{a-1}^2\{\sqrt{n}\tanh(\phi)\} \right.
\nonumber \\&& -\left.\frac{1}{2\sqrt{n}a}U_{2a-1}\{\sqrt{n}\tanh(\phi)\} \right\} \left(1-(n-1)\sinh^2(\phi)\right),
\label{eom1}
\een
where $U_{b}\{\sigma\}$ is defined by Eq.~\eqref{Chebyshev}. 

\begin{figure}
\includegraphics[width=7.5cm,height=6cm]{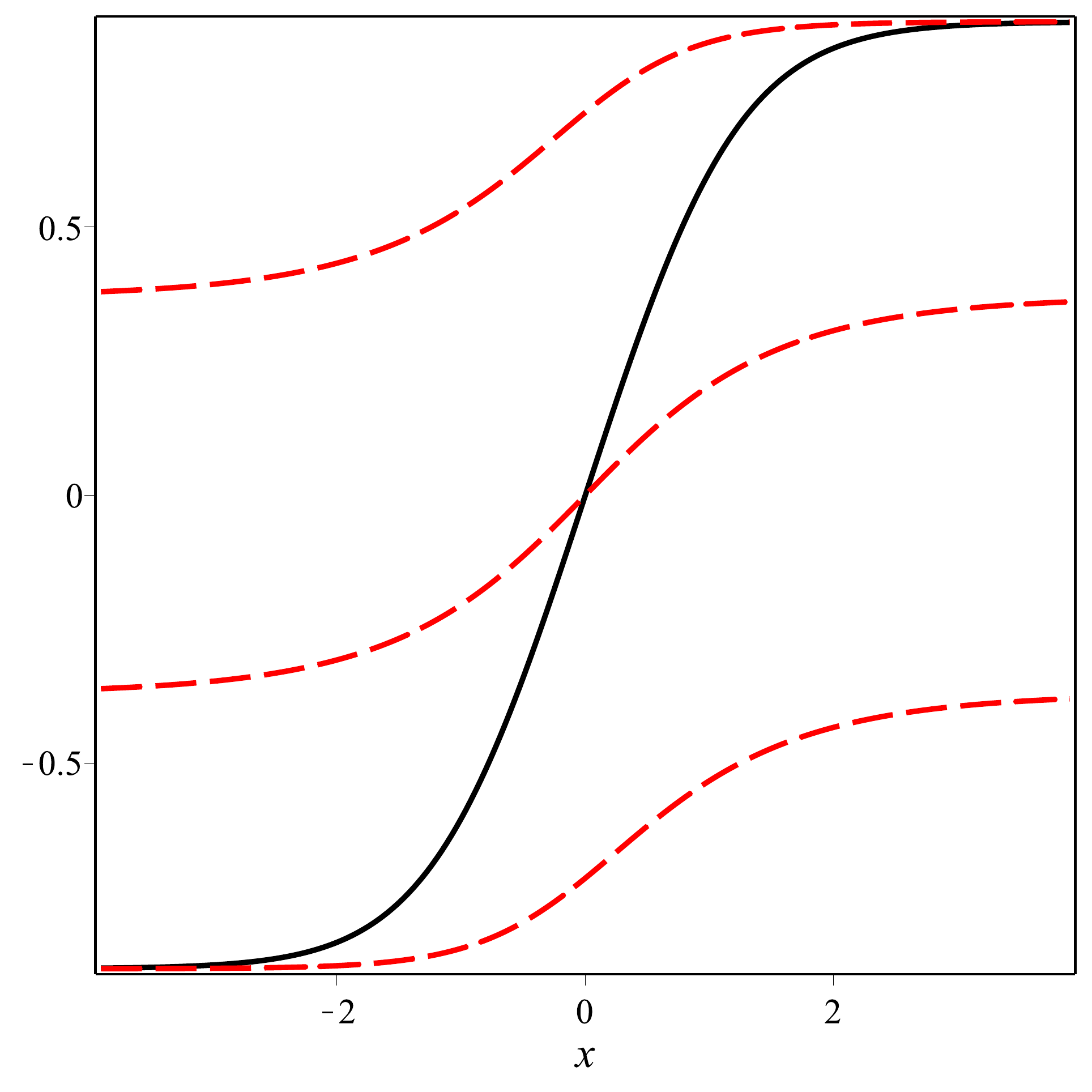}
\includegraphics[width=7.5cm,height=6cm]{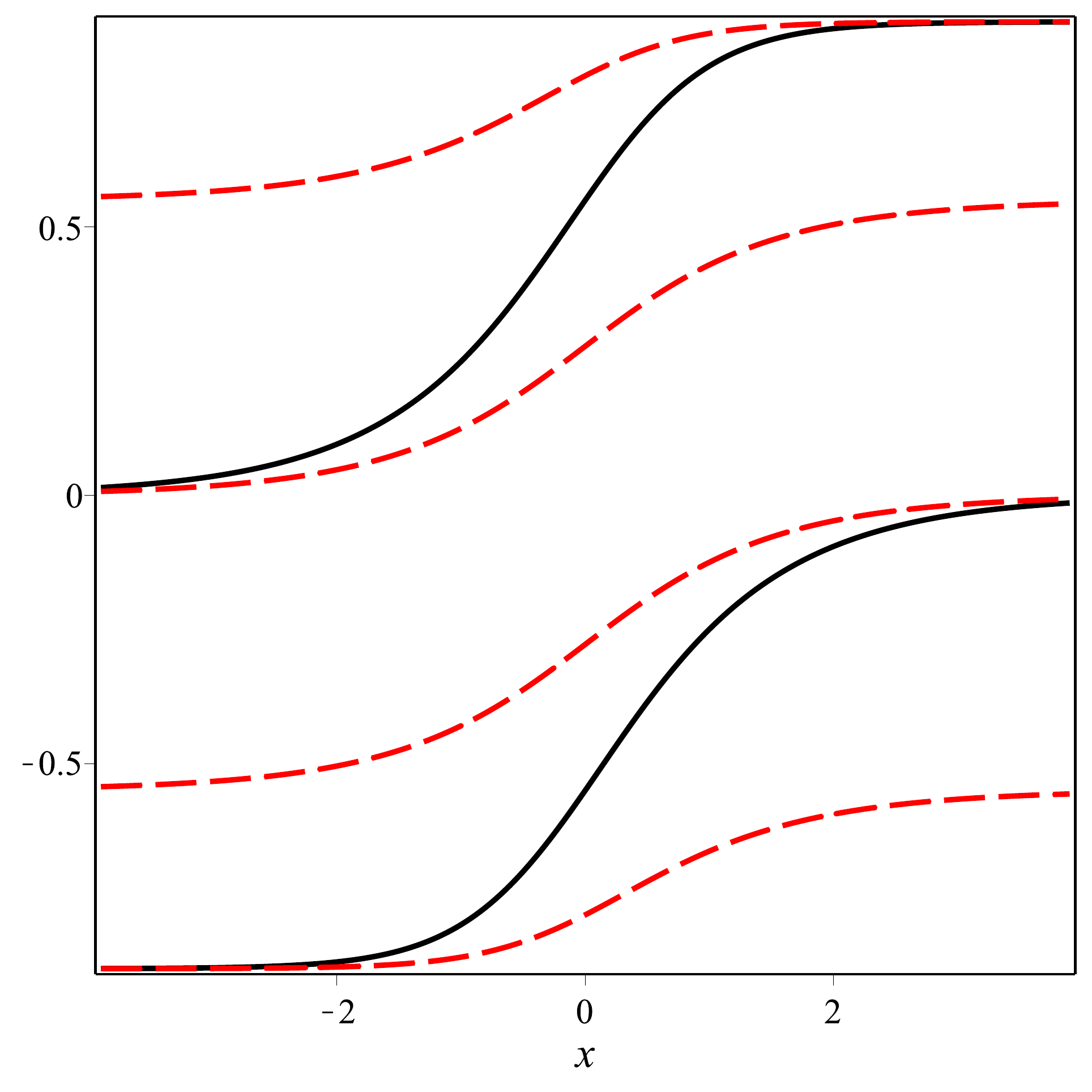}
\caption{The topological solutions \eqref{sol} for $n=2$. In the left panel one shows the case $a=1\, (m=0) $  depicted with solid (black), and $a=3\, (m=0,1,2)$ represented by dashed (red) lines. In the right panel, one takes $a=2\, (m=0,1)$ depicted with solid (black), and $a=4\, (m=0,1,2,3)$ represented by dashed (red) lines. }
\label{fig:2}
\end{figure}

Since the static solutions of \eqref{chi4} are known, we are then able to get all the solutions of the Eq.~\eqref{eom1} connecting distinct minima of the potential $V_a^{(1)}(\phi)$. From the inverse of the deformation function \eqref{def1}, we find
\be
\label{sol}
\phi(x)=\arctanh\left\{\frac{1}{\sqrt{n}}\cos\left(\frac{\theta(x)+m\pi}{a}\right)\right\},
\ee
where $\theta(x)=\arccos(\tanh(x))$ and $m=0,1,...,2a-1$. Each value of $m$ produces solution connecting  different topological sectors. Moreover, for $m=0,..,a-1$ one gets kink solutions, and for $m=a,..,2a-1$ one gets antikink solutions. For purposes of illustration, we display in the Fig.~\ref{fig:2} the kinklike solutions for the potentials represented in the Fig.~\ref{fig:1}.

The energy density of the solutions are
\be
\label{en1}
\rho^{(1)}_{a}(x)=\frac{n}{a^2}\frac{\sin^2\left(\frac{\theta(x)+m\pi}{a}\right)\sech^2(x)}{\left[n-\cos^2\left(\frac{\theta(x)+m\pi}{a}\right)\right]^2}.
\ee

We also study the linear stability, as seen in Eq.~\eqref{pet}, of the model under consideration and we find the following stability potential 
\ben
U_{a}^{(1)}(x)&=&1-2\sech^2(x) + \frac{3}{a}\sech(x)\tanh(x)
 \frac{\cotg\left(\frac{\theta(x)+m\pi}{a}\right)\left[n-2+\cos^2\left(\frac{\theta(x)+m\pi}{a}\right)\right]}{n-\cos^2\left(\frac{\theta(x)+m\pi}{a}\right)} \nonumber \\ & & 
-\frac{\sech^2(x)}{na^2}\left\{n-2
+2\cos^2\left(\frac{\theta(x)+m\pi}{a}\right) \frac{\left[n(4n-5)+\cos^2\left(\frac{\theta(x)+m\pi}{a}\right)\right]} {\left(n-\cos^2\left(\frac{\theta(x)+m\pi}{a}\right)\right)^2} \right\}.\nonumber \\
\een
The zero mode \eqref{zm} is represented by
\be
\eta_{0,a}^{(1)}(x)=\frac{\sqrt{n}}{a}\frac{\sin\left(\frac{\theta(x)+m\pi}{a}\right)\sech(x)}{\left[n-\cos^2\left(\frac{\theta(x)+m\pi}{a}\right)\right]} .
\ee

\begin{figure}
\centering
\includegraphics[width=7cm,height=6.5cm]{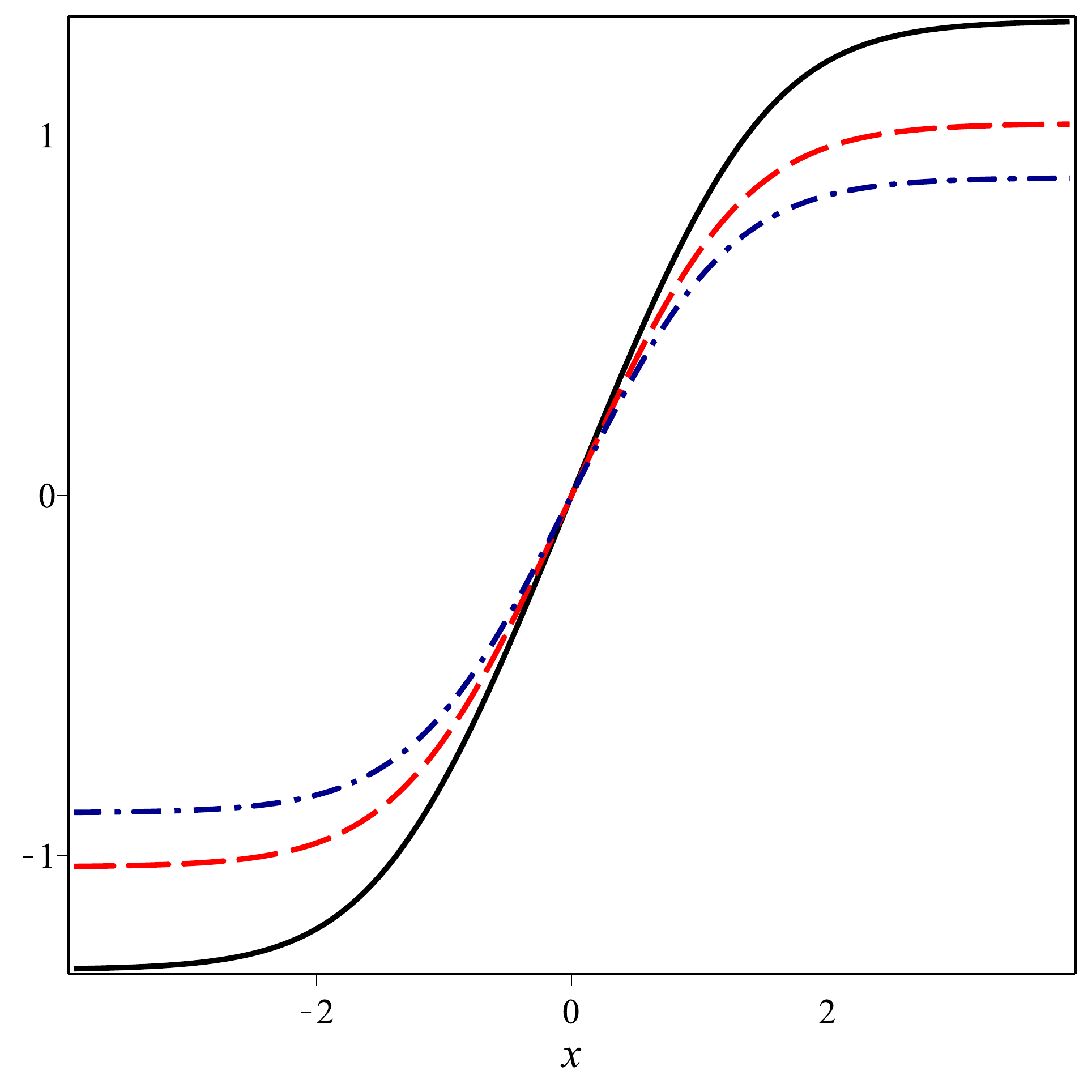}
\includegraphics[width=7cm,height=6.5cm]{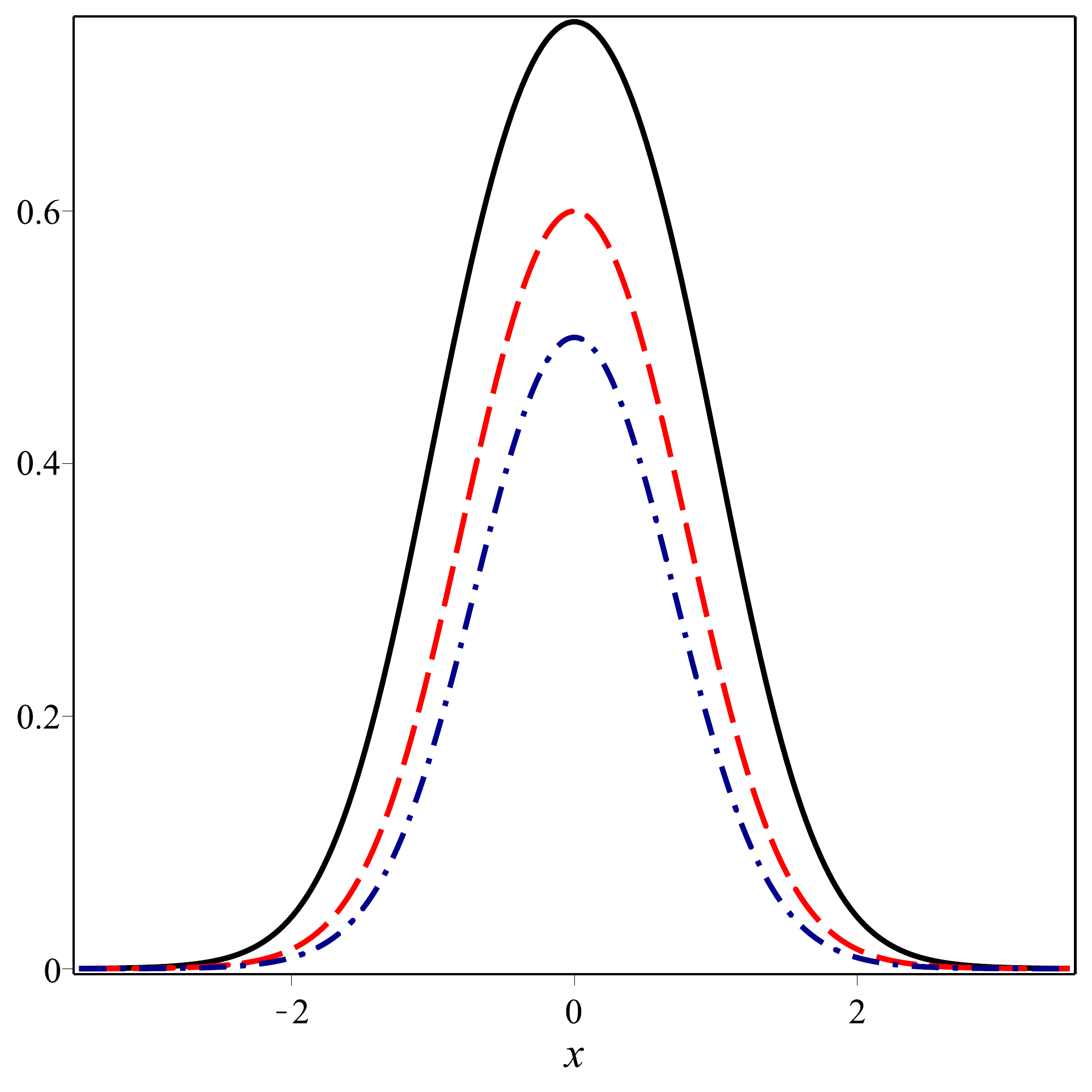}
\includegraphics[width=7cm,height=6.5cm]{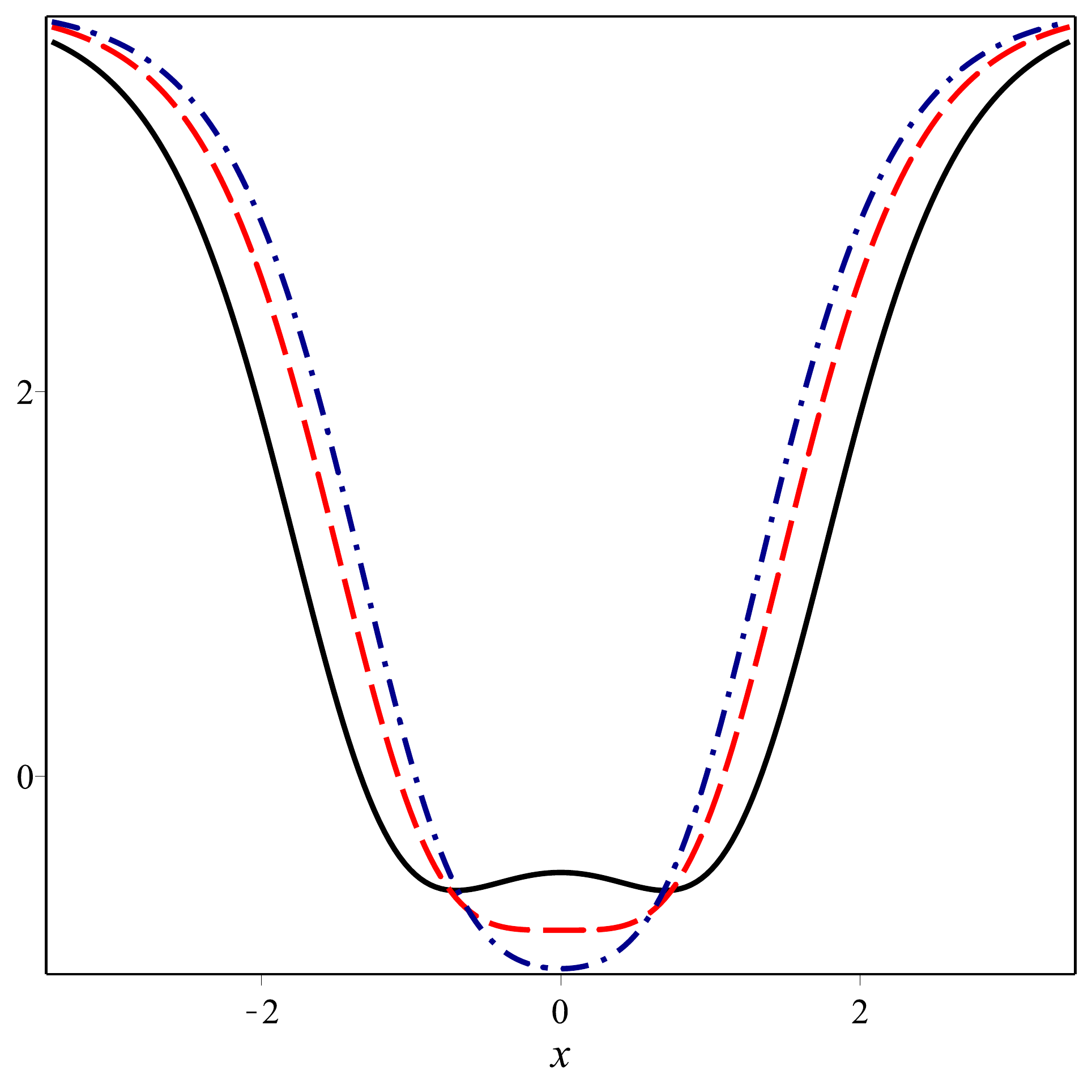}
\includegraphics[width=7cm,height=6.5cm]{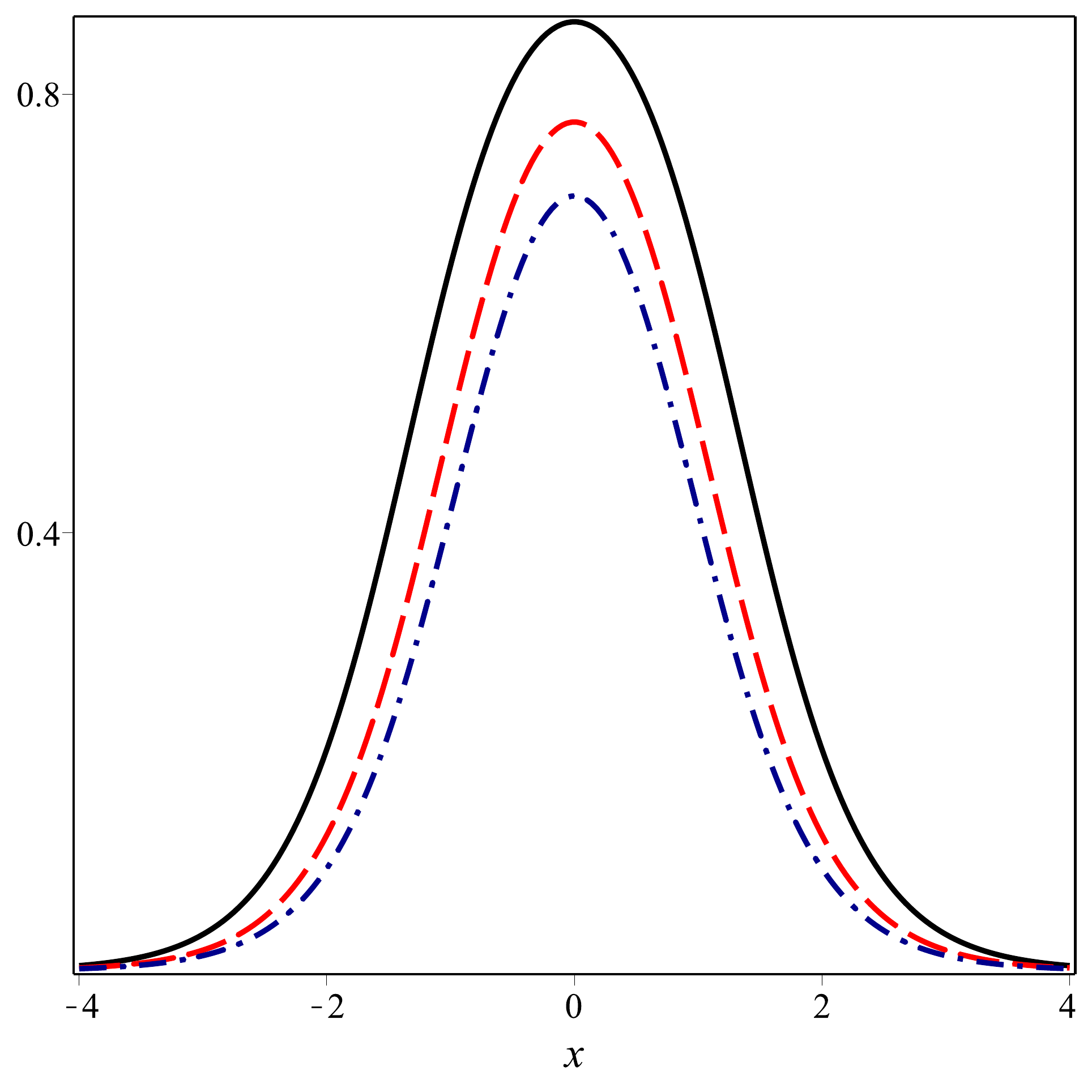}
\caption{In the top panel, one shows the kink solutions \eqref{psa1} (left) and their energy densities \eqref{rsa1} (right). In the bottom panel, one depicts  the stability potentials \eqref{stab1} (left) and the corresponding zero modes \eqref{mzero1} (right). We vary $n$ as $n=4/3,5/3,2$, represented by solid (black), dashed (red) and dot-dashed (blue) lines, respectively.}
\label{fig:3}
\end{figure}

Particularly, we  intend to highlight properties of a system described by this new potential when $a=1$.  In this standard scenario, the kink solution, its energy density, the stability potential and zero mode are explicitly given by
\ben
\label{psa1}
\phi^{(1)}_{1}(x)&=&\arctanh\left[ \frac{1}{\sqrt{n}} \tanh(x) \right]; \\
\label{rsa1}
\rho^{(1)}_{1}(x)&=& \frac{n}{\left[1+(n-1)\cosh^2(x)\right]^2};  \\
\label{stab1}
U_{1}^{(1)}(x)&=&\frac{2(n-1)\left[2(n-1)\cosh^4(x)- (3n-1)\cosh^2(x)+1\right]}{\left[1+(n-1)\cosh^2(x)\right]^2}; \\
\label{mzero1}
\eta_{0,1}^{(1)}(x)&=&\frac{\sqrt{n}}{1+(n-1)\cosh^2(x)}. 
\een
In Fig. \ref{fig:3}, these quantities are displayed for some values of $n$. It is shown that the  stability potential is of the modified P$\ddot{\o}$schl-Teller type \cite{PT1,PT2}, with at least one bound state, the zero mode. We notice that the stability potential has a peculiar behavior near the origin, where a transition from a maximum to a minimum occurs when $n=5/3$.

\subsection{Second family of models}
\begin{figure}
\includegraphics[width=7.5cm,height=6cm]{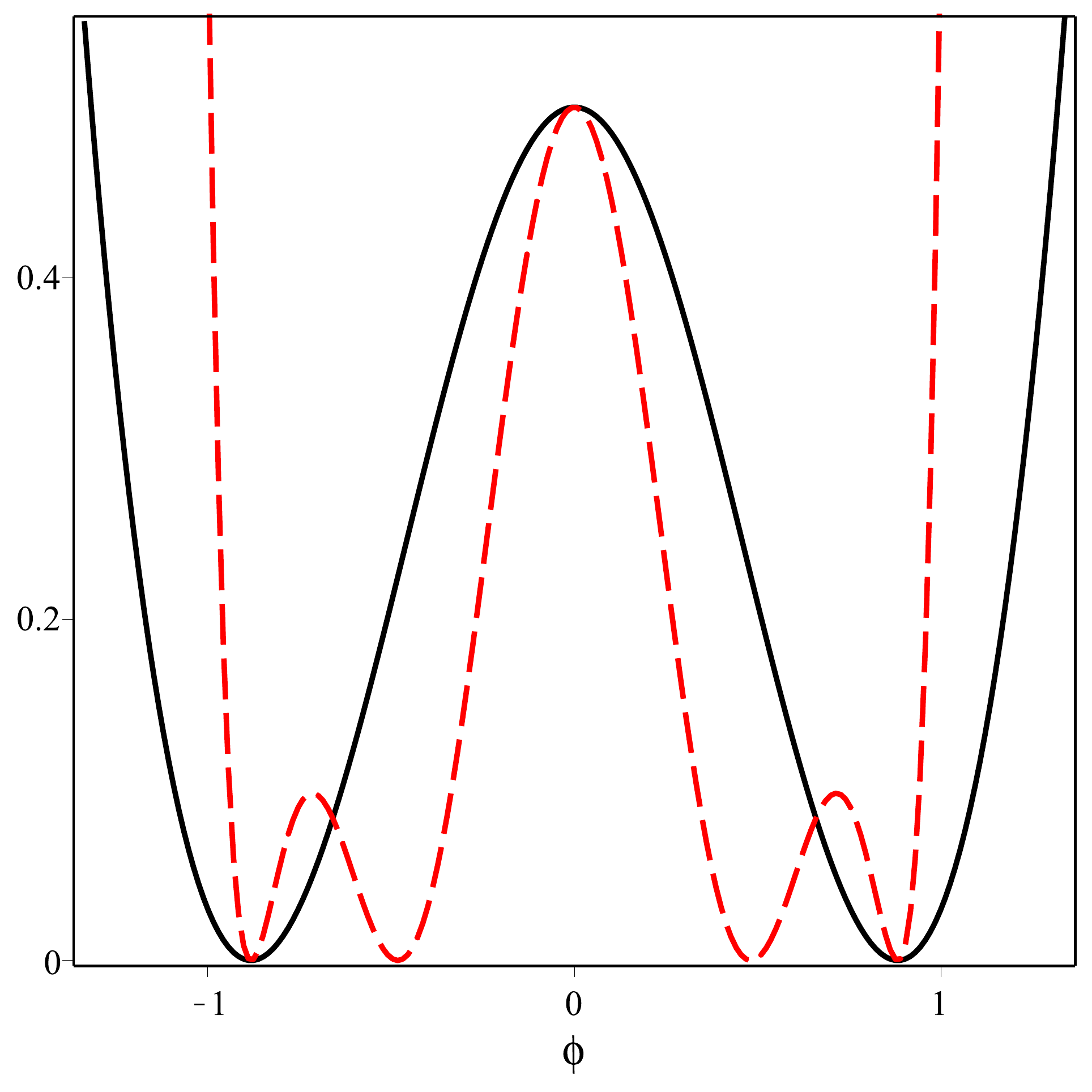}
\includegraphics[width=7.5cm,height=6cm]{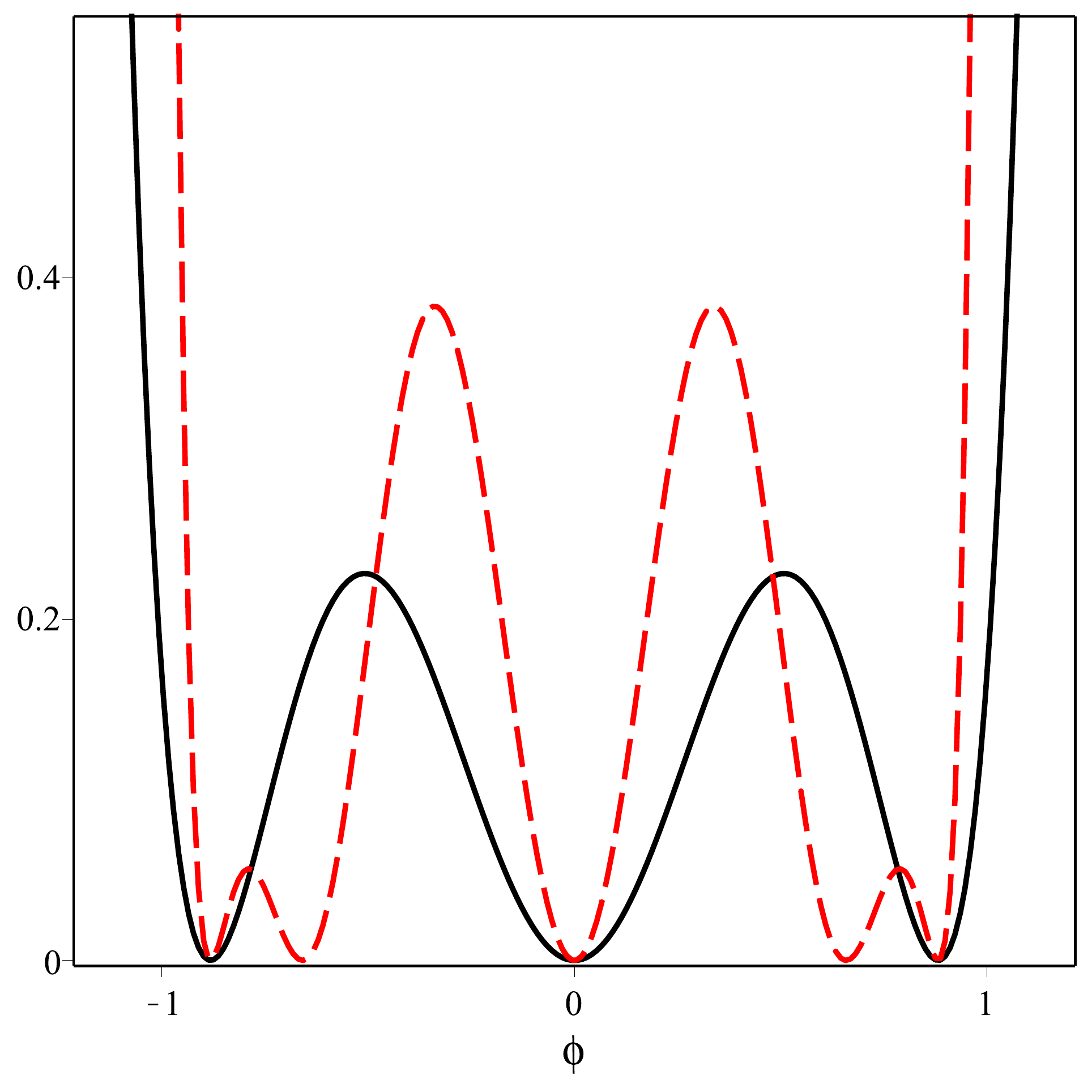}
\caption{The potentials \eqref{pott2}, depicted as $a^2V_a^{(2)}(\phi)$ for $n=1$. In the left panel, one plots $V_1^{(2)}(\phi)$ and
$V_3^{(2)}(\phi)$ with solid (black) and dashed (red) lines, respectively. In the right panel, one shows $V_2^{(2)}(\phi)$ and $V_4^{(2)}(\phi)$, depicted with solid (black) and dashed (red) lines, respectively.}
\label{fig:4}
\end{figure}

Following the same procedure, we now choose another deformation function
\be
f^{(2)}(\phi)=\cos\left[a \,\arccos\left(\sqrt{n}\,\sinh(\phi)\right)-m\pi\right]\,,
\ee
where the constants $a$ and $m$ have the same values used before, but now $n$ is positive real. The deformed potential can be written in the form
\be
V_a^{(2)}(\phi)=\frac{\sech^2(\phi)}{2na^2}\left(1-n\,\sinh^2(\phi)\right)\left(1-[f^{(2)}(\phi)]^2\right),
\ee
which can also be written in terms of the Chebyshev polynomials
\be
\label{pott2}
V_a^{(2)}(\phi)=\frac{1}{2na^2}\frac{(1-n\,\sinh^2(\phi))^2}{\cosh^2(\phi)}U_{a-1}^{2}\{\sqrt{n}\,\sinh(\phi)\}\,.\\
\ee
Explicit results of $V_a^{(2)}(\phi)$ for $a= 1, 2, 3, 4$ are given by 
\ben\label{vf2}
V_1^{(2)}(\phi)&=&\frac{\sech^2(\phi)}{2n}(1-n\,\sinh^2(\phi))^2; \\
V_2^{(2)}(\phi) &=&\frac{\tanh^2(\phi)}{2}(1-n\,\sinh^2(\phi))^2;\\
V_3^{(2)}(\phi)&=&\frac{{\rm sech}^2(\phi)}{18n}(1-4n\,\sinh^2(\phi))^2(1-n\,\sinh^2(\phi))^2;\\
V_4^{(2)}(\phi)&=&\frac{\tanh^2(\phi)}{2}(1-2n\,\sinh^2(\phi))^2(1-n\,\sinh^2(\phi))^2.
\een
These potentials are displayed in the Fig.~\ref{fig:4}.

\begin{figure}
\includegraphics[width=7.5cm,height=6cm]{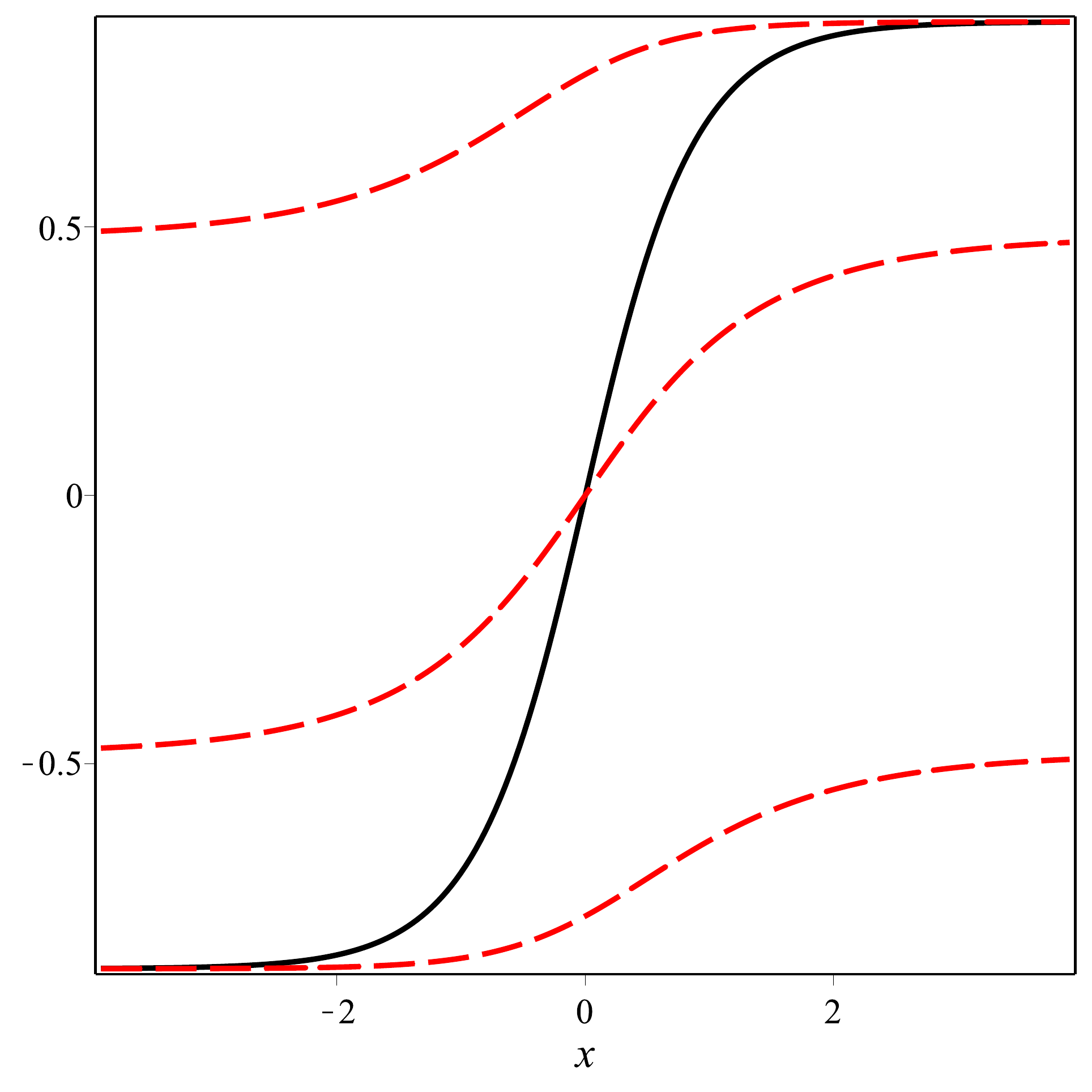}
\includegraphics[width=7.5cm,height=6cm]{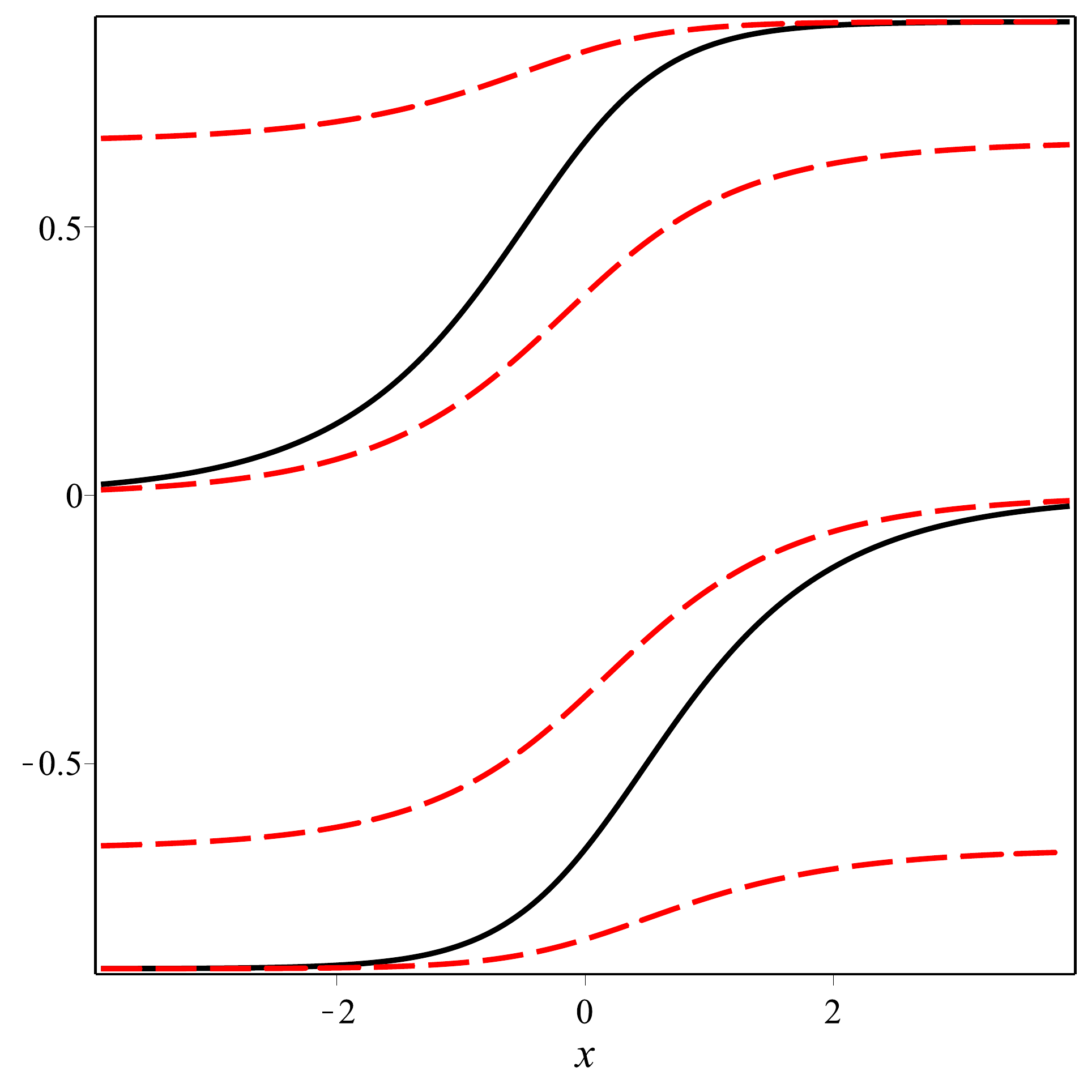}
\caption{The topological solutions \eqref{sol2} for $n=1$. In the left panel, one takes $a=1 \, (m=0) $  depicted with solid (black) and
$a=3 \, (m=0,1,2)$ represented by dashed (red) lines. In the right panel, one shows $a=2 \, (m=0,1)$ depicted with solid (black), and
$a=4 \, (m=0,1,2,3)$ represented by dashed (red) lines. }
\label{fig:5}
\end{figure}

The minima of the potentials are given by
\be
\bar{\phi}=\arcsinh\left[\frac{1}{\sqrt{n}}\cos\left(\frac{k\pi}{a}\right)\right]
\ee
where $k=0,..,a$. Admitting static fields, the equation of motion is given by
\ben
\frac{d^2\phi}{d x^2}&=&-\left(1-n\sinh^2(\phi)\right)\left\{\frac{\sech(\phi)}{2\sqrt{n}a}U_{2a-1}\{\sqrt{n}\sinh(\phi)\} \right. \nonumber \\
&&  +\left.\frac{(n+1)}{na^2}\tanh(\phi) \sech^2(\phi)U_{a-1}^2\{\sqrt{n}\sinh(\phi)\}\right\}.
\label{eom2}
\een
The set of solutions that satisfies the equation of motion \eqref{eom2} is
\be
\label{sol2}
\phi(x)=\arcsinh\left\{\frac{1}{\sqrt{n}}\cos\left(\frac{\theta(x)+m\pi}{a}\right)\right\},
\ee
where $\theta(x)=\arccos(\tanh(x))$. The kink solutions are found for $m=0,..,a-1$, and the antikinks for  $m=a,..,2a-1$. We plot in the Fig. \ref{fig:5} the kinklike solutions for the potentials represented in the Fig.~\ref{fig:4}.

\begin{figure}
\centering
\includegraphics[width=7cm,height=6.5cm]{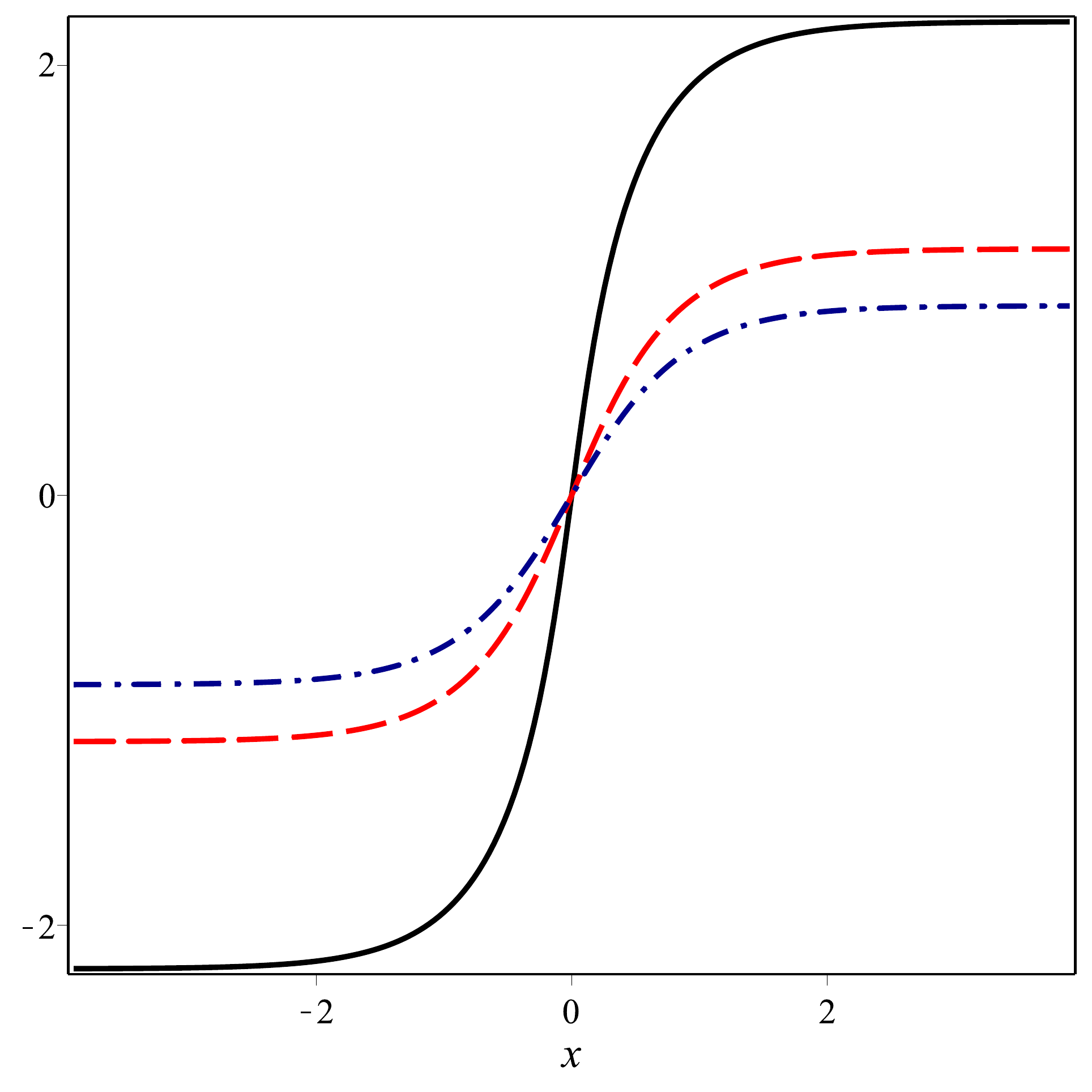} 
\includegraphics[width=7cm,height=6.5cm]{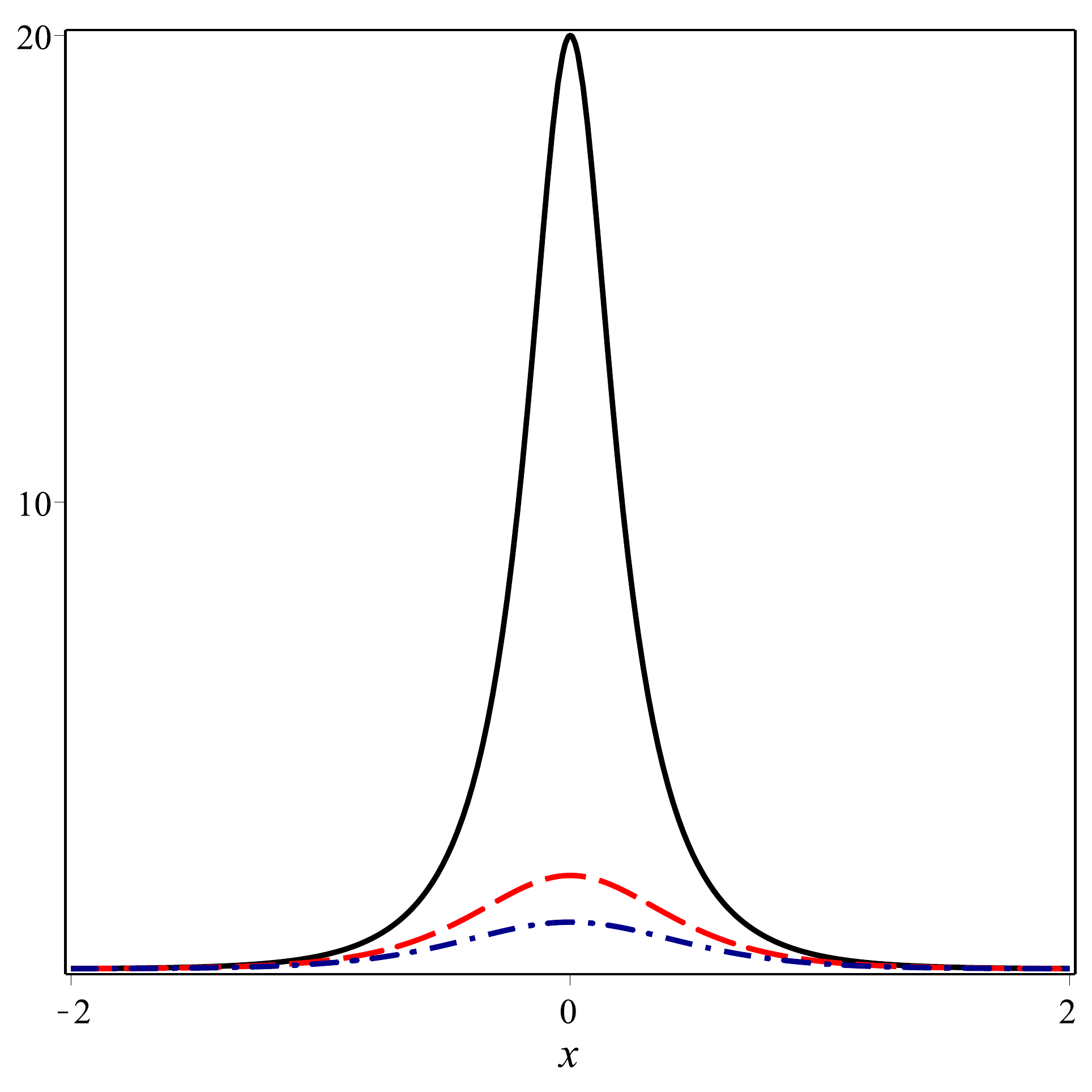}  
\includegraphics[width=7cm,height=6.5cm]{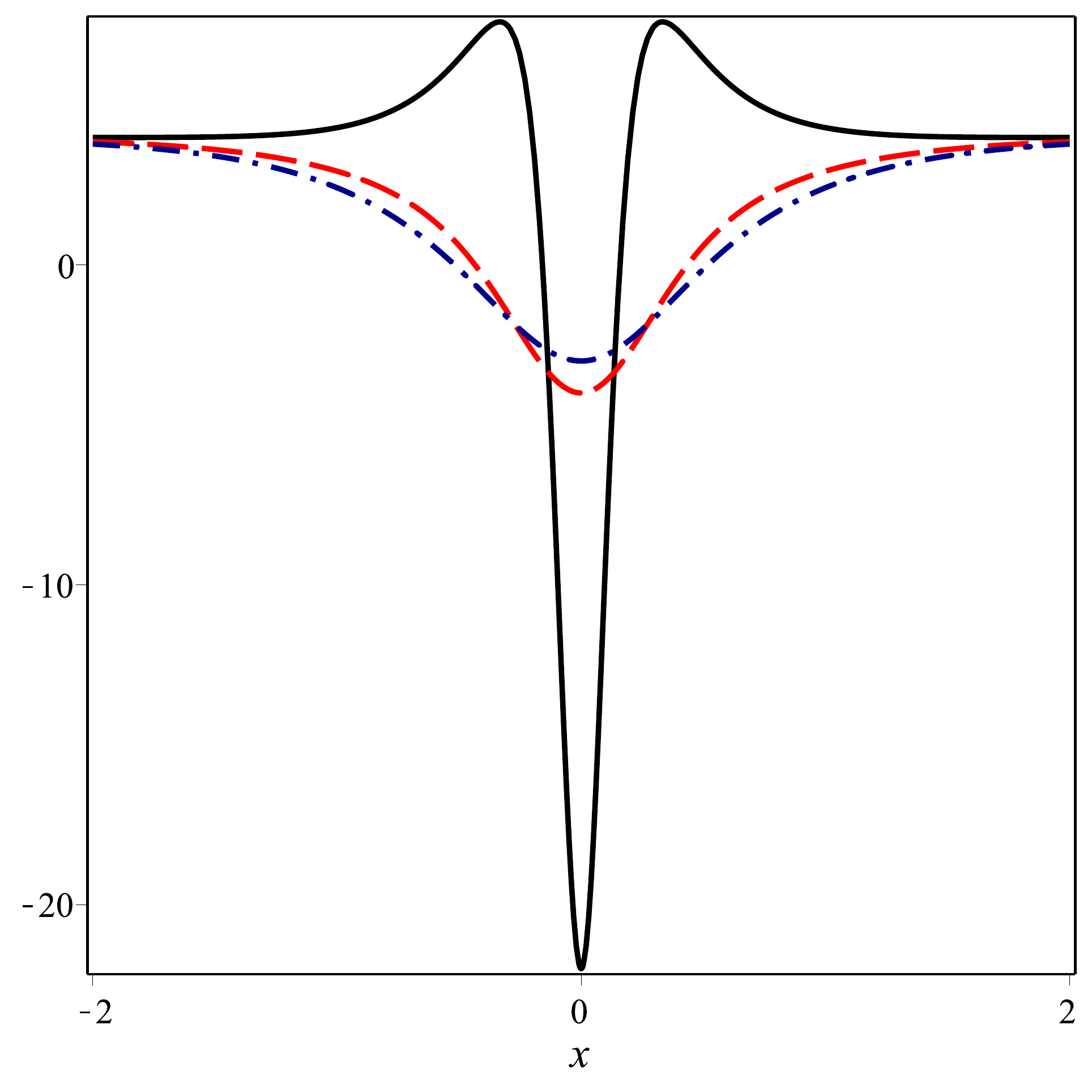}
\includegraphics[width=7cm,height=6.5cm]{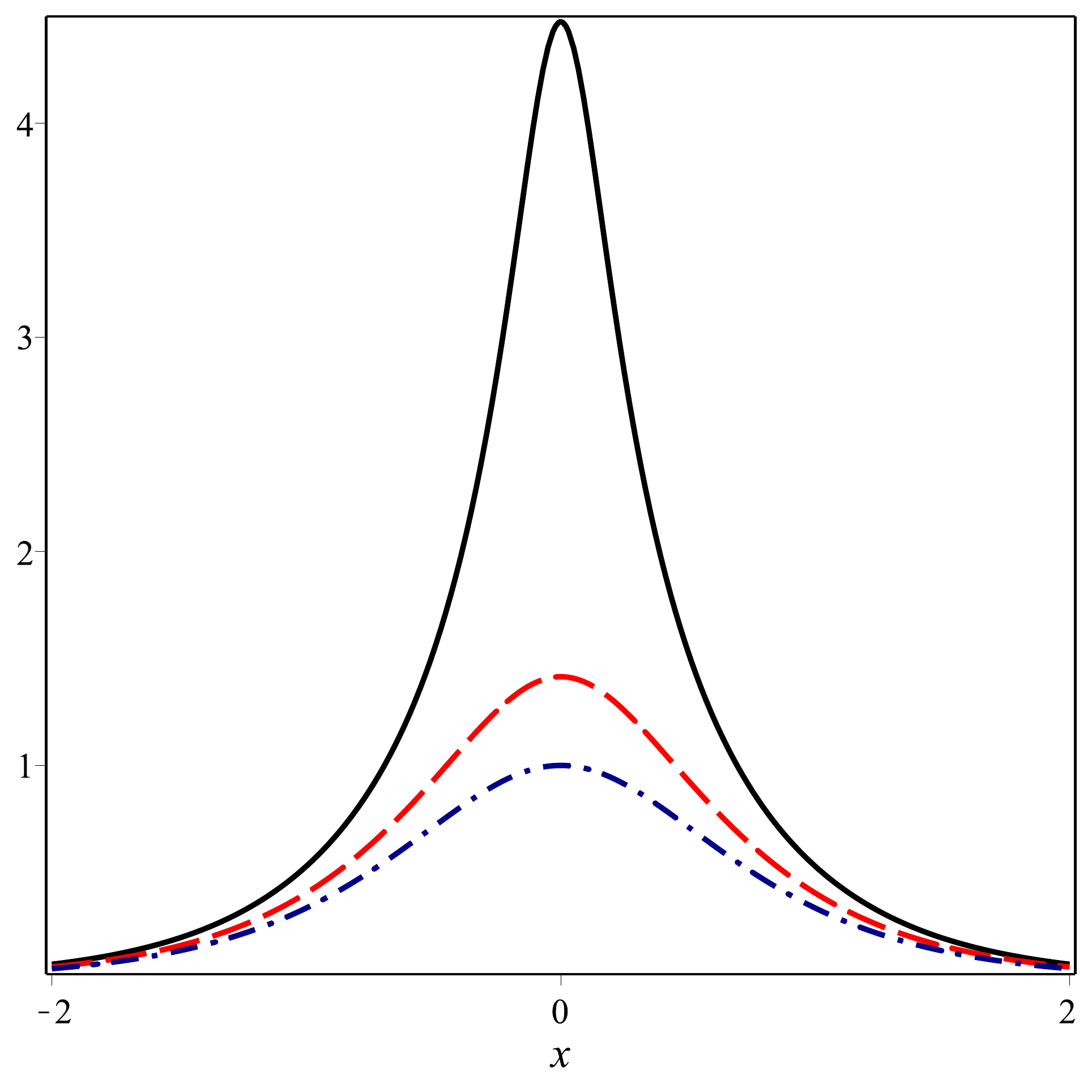}
\caption{In the top panel one displays the kink solution \eqref{sol2} (left) and its energy density \eqref{rho2} (right), for $a=1$. In the bottom panel one shows the stability potential \eqref{stab2} (left) and the corresponding zero mode \eqref{mzero2} (right). We are using $n=0.05,1/2,1$, represented by solid (black), dashed (red) and dot-dashed (blue) lines, respectively.} 
\label{fig:6}  
\end{figure}

The energy densities of the solutions are given by
\be
\label{rho2}
\rho^{(2)}_{a}(x)=\frac{1}{a^2}\frac{\sin^2\left(\frac{\theta(x)+m\pi}{a}\right)\sech^2(x)}{\left[n+\cos^2\left(\frac{\theta(x)+m\pi}{a}\right)\right]}.
\ee
The stability potential is given as follows
\ben
U_{a}^{(2)}(x)&=&1-2\sech^2(x) 
+\frac{3}{a}(n+1)\sech(x)\tanh(x)\frac{\cotg\left(\frac{\theta(x)+m\pi}{a}\right)}{n+\cos^2\left(\frac{\theta(x)+m\pi}{a}\right)} \nonumber\\ & &
+\frac{1}{a^2}(n+1)\sech(x)^2\frac{\left[2\cos^2\left(\frac{\theta(x)+m\pi}{a}\right)-n\right]}{\left(n+\cos^2\left(\frac{\theta(x)+m\pi}{a}\right)\right)^2}. 
\een
And the zero mode is represented by
\be
\eta_{0,a}^{(2)}(x)=\frac{\sin\left(\frac{\theta(x)+m\pi}{a}\right)\sech(x)}{a\sqrt{n+\cos^2\left(\frac{\theta(x)+m\pi}{a}\right)}}.
\ee
In special, for $a=1$ we have 
\ben
\label{stab2}
U_{1}^{(2)}(x)&=&\frac{4(n+1)^2\cosh^6(x)-(n+1)(6n+8)\cosh^4(x)+(5n+6)\cosh^2(x)-2}{\cosh^2(x)\left[(n+1)\cosh^2(x)-1\right]^2}\:\:\:\:\:\:\:\:\:\:\:\:\:\:\:\:\:\:\:\:\:\: \\
\label{mzero2}
\eta_{0,1}^{(2)}(x)&=&\frac{\sech^2(x)}{\sqrt{n+\tanh^2(x)}}.
\een
%
These quantities  are shown in the Fig. \ref{fig:6} for some values of $n$,  together with the kink solution and its energy density.   Here the stability potential  is of volcano type for small $n$ and it changes its behavior to a modified P$\ddot{\o}$schl-Teller type \cite{PT1,PT2} as $n$ increases. Potentials with such properties were also found very recently in Ref.~\cite{bemfica}.

\section{Generalized models} \label{sec-3}

Let us now consider such models in a non-standard perspective, in which one introduces a modification of the kinematic term in the Lagrange density, that is, one changes \eqref{lagran} to
\be\label{Lg1}
{\mathcal L}=-\frac{1}{4}\left(\partial_\mu \phi \partial^\mu \phi\right)^2-V(\phi).
\ee 
For static configurations, the equation of motion and the energy density are given by
\bens
\slabel{soe}
3\phi'^2\phi''&=&V_{\phi},\\
\slabel{esoe}
\rho(x)&=&\frac{1}{4}\phi'^4+V(\phi),
\eens
where primes denote differentiation with respect to the variable $x$, and $V_{\phi}=dV/d\phi$. In addition, we bring to the current investigation the description of the first order formalism shown in Ref.~\cite{FOGD}. In this context, if we can  write 
\be
V(\phi)=\frac{3}{4}W_{\phi}^{4/3},
\ee
the first order equation
\be
\phi'=W_{\phi}^{1/3}
\ee
satisfies the second order equation of motion and the energy density acquires a simpler form
\be
\rho(x)=W_{\phi}^{4/3}=\phi'^4.
\ee

The stability potential is calculated as shown in Ref.~\cite{FOFE}. The procedure is similar to the case studied before, and for
$\phi(x,t)= \phi(x) +\sum_n\,\eta_{n}(x)\cos(\omega_{n} t)$ one gets to a Schr$\ddot{\o}$dinger-like equation
\be\label{srole}
\left(-\frac{d^2}{dz^2}+U(z)\right)u_n=\omega_n^2u_n,
\ee
where
\be
\label{epof}
U(z)=\frac{1}{\phi_{z}^2}\left(3V_{\phi\phi}+\phi_{z}\phi_{zzz}\right),
\ee 
and we have made the following changes:
\be
x=\sqrt{3} z \:\:\: \:\:\: \mbox{and} \:\:\: \:\:\: \eta_n=\frac{u_n}{3^{1/4}\phi'}.
\ee 
Considering the first order formalism \cite{FOGD},
\be
U(z)=2W_{\phi}^{-1/3}W_{\phi\phi\phi},
\ee
it is possible to write \eqref{srole} in a factorizable manner  
\be
A^{\dagger}A u_n=\left(-\frac{d^2}{dz^2}+2W_{\phi}^{-1/3}W_{\phi\phi\phi}\right)u_n=\omega_n^2u_n,
\ee
where
\ben
A&=&-\frac{d}{dz}+\frac{2}{\sqrt{3}}W_{\phi}^{-2/3}W_{\phi\phi}, \nonumber\\
\een
which leads to non-negative eigenvalues $\omega_n^2$  and therefore to stable solutions. 

\subsection{First model}

As can be seen, defect structures with modification in the kinematics introduces new kinds of non-linearities, whose study is of interest to diverse areas of non-linear science. Here we will show that it is possible to get analytical results for the specific potential \eqref{vf1} under these circumstances. For convenience, we rewrite that potential as
\be\label{vcomp1}
V(\phi)=\frac{3}{4}\left(1-\alpha^2\sinh^2(\phi)\right)^2\,,
\ee
where $\alpha$ is a real number. This model has two minima at $\bar{\phi}_{\pm}=\pm \arcsinh(1/\alpha)$.
In this case,  
\ben \label{e2oo}
\phi'^2\phi''&=&-\frac12\alpha^2\sinh(2\phi)\left(1-\alpha^2\sinh^2(\phi)\right)\,,\\
\rho(x)&=&\frac{1}{4}\phi'^4+\frac{3}{4}\left(1-\alpha^2\sinh^2(\phi)\right)^2.
\een
One follows the first order formalism to see that the Eq.\eqref{e2oo} is reduced to
\be\
\phi^{\prime}=\sqrt{1-\alpha^2\sinh^2(\phi)}, 
\ee
whose solution is compact-like, given by
\begin{eqnarray}\label{comp1}
\phi(x) = \left\{
\begin{array}{ll}
-\arcsinh(1/{\alpha}), \, x<-\bar{x},\\ 
 \arcsinh\left[\dfrac{1}{\alpha}\sn(\alpha x,{i}/{\alpha})\right], \, |x|\leq\bar{x},\\
\arcsinh(1/{\alpha}), \, x>\bar{x}\,,
\end{array} \right.
\end{eqnarray}
where $\sn(\alpha x,{i}/{\alpha})$ is a Jacobi elliptic function with argument $\alpha x$ and modulus ${i}/{\alpha}$  as variables,  $i=\sqrt{-1}$, and $\alpha\,\bar{x}=K({i}/{\alpha})$ with $K({i}/{\alpha})$ being a complete elliptic integral of the first kind. Alternatively, we can use the transformations for elliptic functions with imaginary modulus \cite{HBEF} to get 
\ben
\frac{1}{\alpha}\sn\left(\alpha x,\frac{i}{\alpha}\right)&=& \frac{1}{\sqrt{1+\alpha^2}}\sd\left(x\sqrt{1+\alpha^2} ,\frac{1}{\sqrt{1+\alpha^2}}\right) \nonumber\\
K\left(\frac{i}{\alpha}\right)&=&\frac{\alpha}{\sqrt{1+\alpha^2}}K\left(\frac{1}{\sqrt{1+\alpha^2}}\right). \nonumber
\een
Consequently, we can rewrite
\begin{eqnarray} \label{comp1a}
\phi(x) =  \left\{
\begin{array}{ll}
-\arcsinh({1}/{\alpha}), \, x<-\bar{x},\\ 
 \arcsinh\left[\dfrac{1}{\sqrt{1+\alpha^2}}\sd\left( x \sqrt{1+\alpha^2} ,\dfrac{1}{\sqrt{1+\alpha^2}}\right)\right], \, |x|\leq\bar{x}, \\
\arcsinh({1}/{\alpha}), \, x>\bar{x}\,.
\end{array} \right.
\end{eqnarray}
\begin{figure}
\centering
\includegraphics[width=7cm,height=6.5cm]{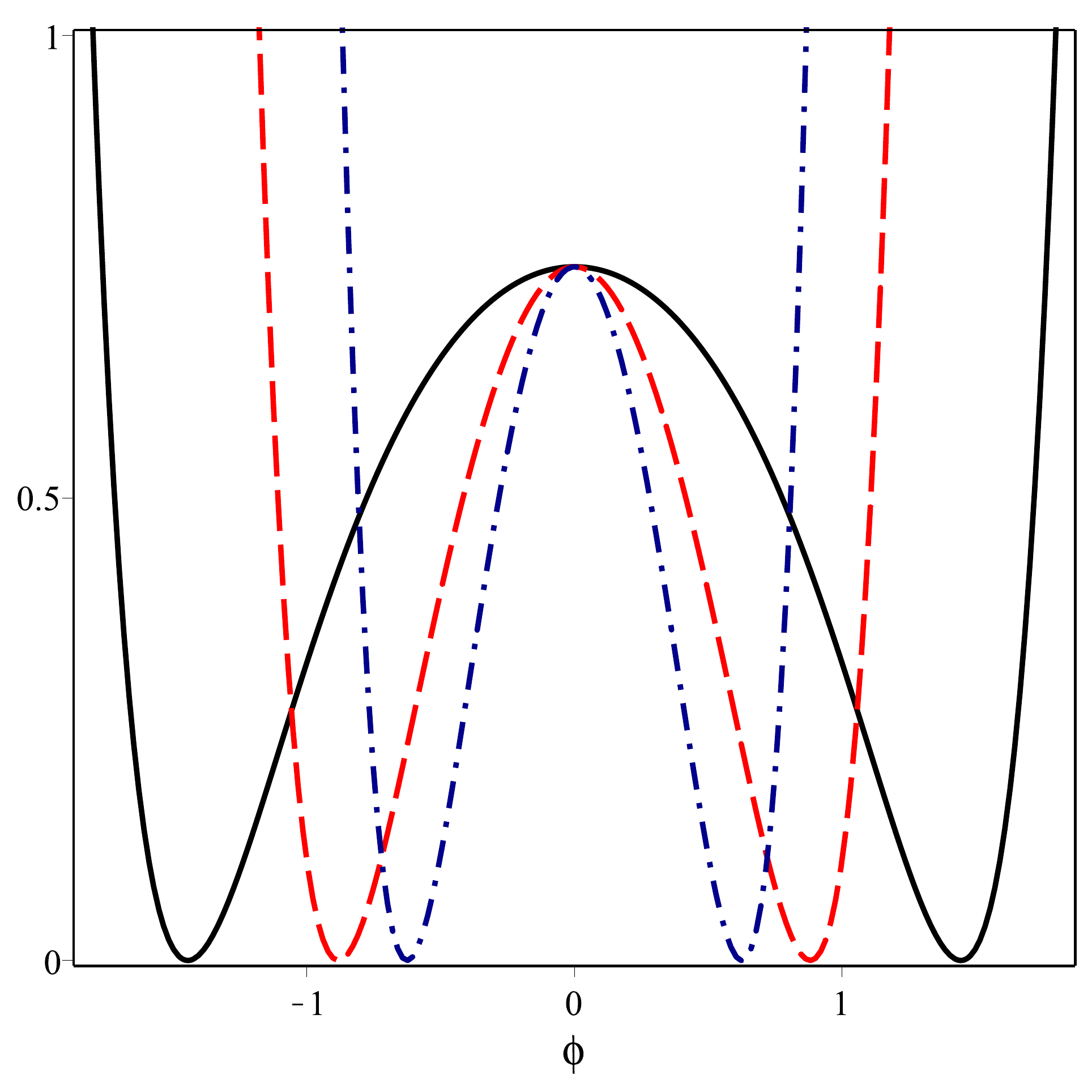}
\includegraphics[width=7cm,height=6.5cm]{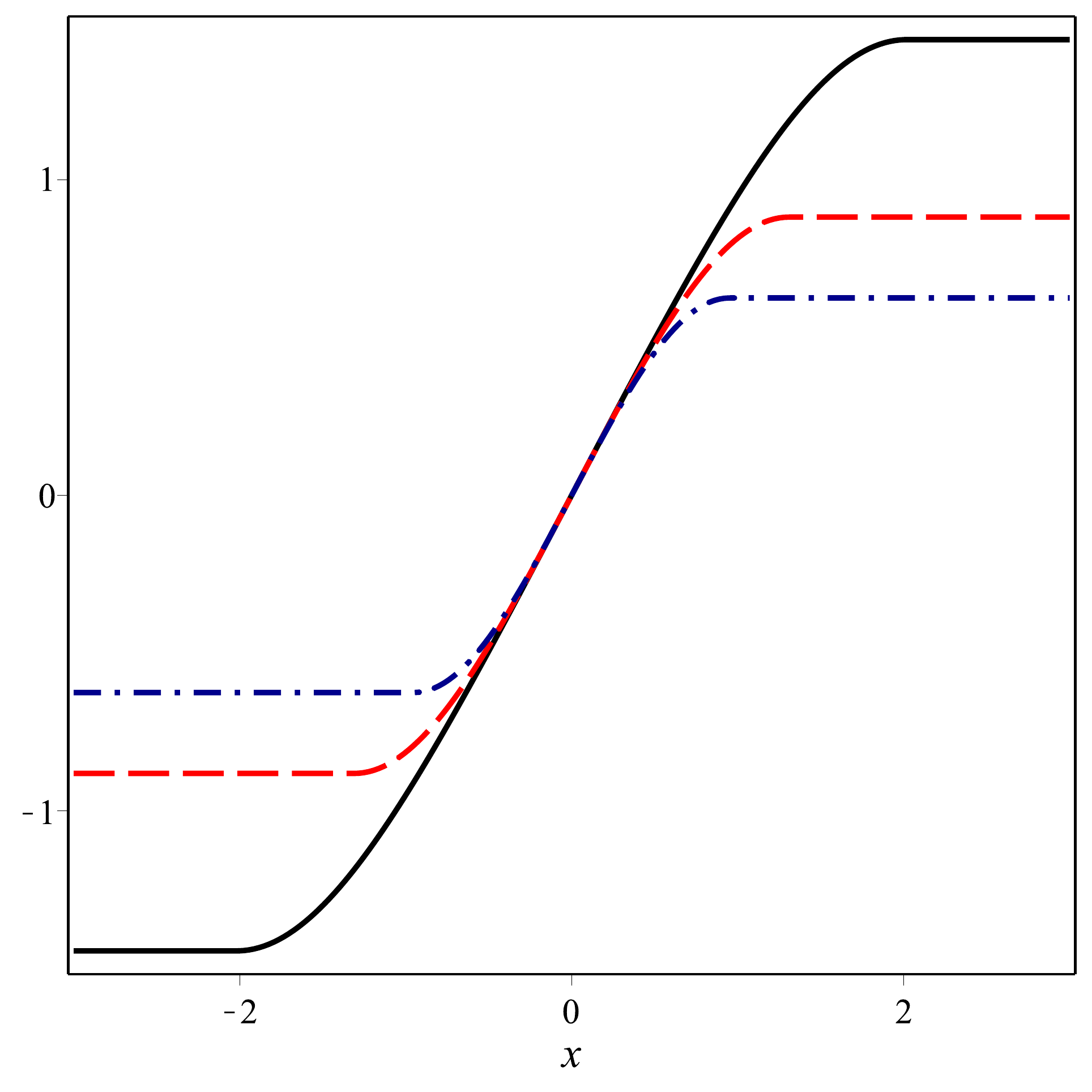}
\includegraphics[width=7cm,height=6.5cm]{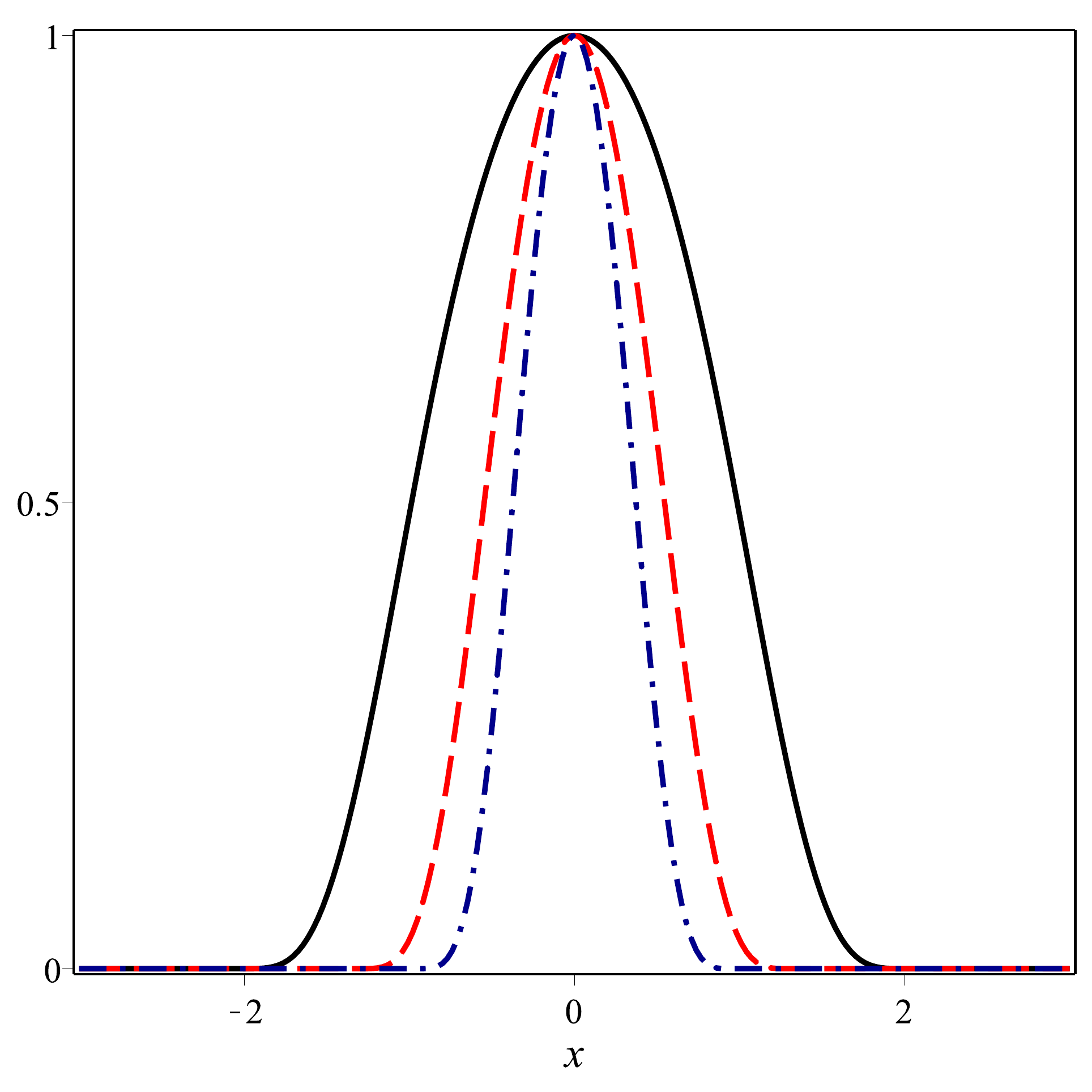}
\includegraphics[width=7cm,height=6.5cm]{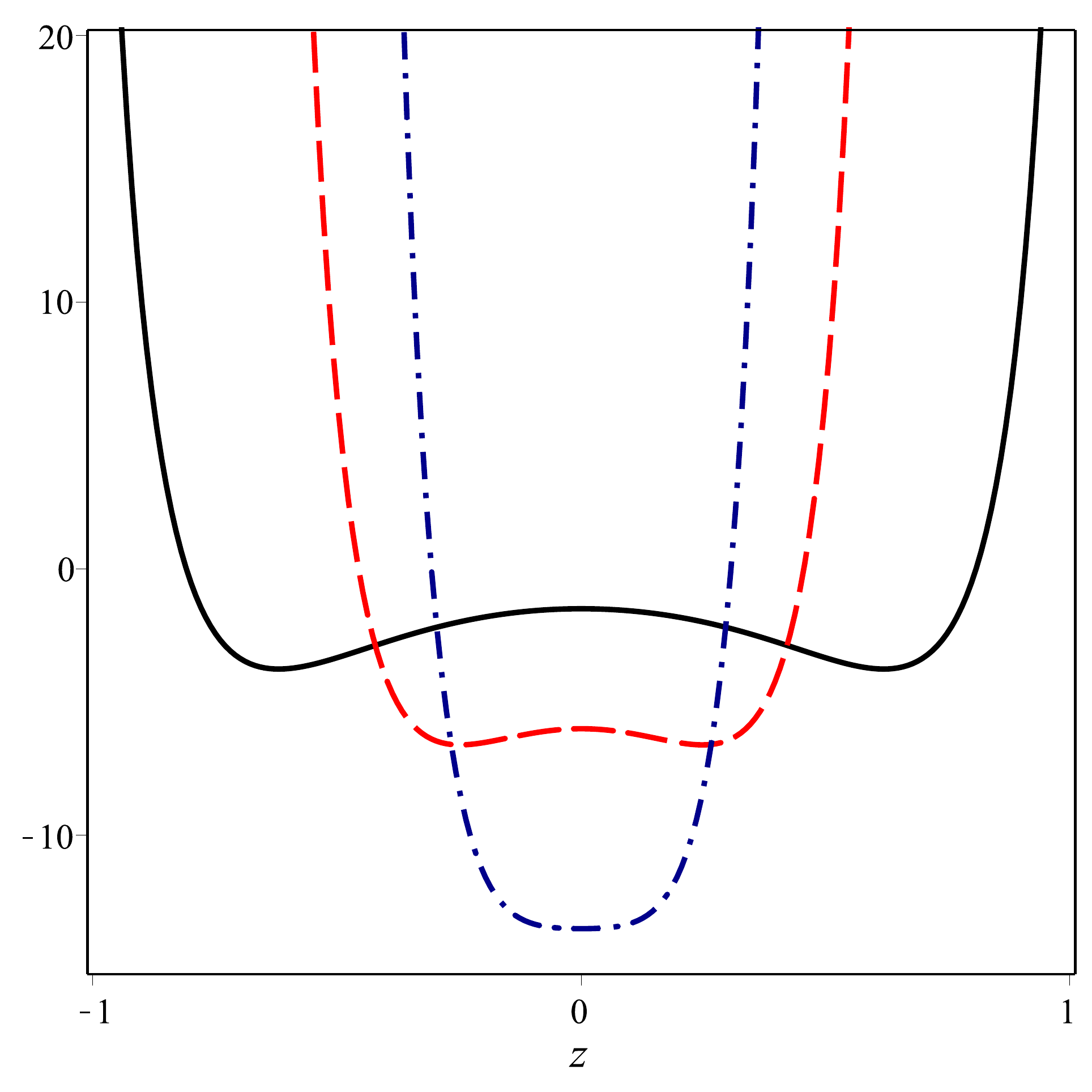}
\caption{In the top panel one shows the potential \eqref{vcomp1} (left) and the compact solution \eqref{comp1a} (right). In the bottom panel one displays the energy density \eqref{ecomp} (left) and the stability potential \eqref{epc} (right). We are using $\alpha=1/2,1,3/2$, represented by solid (black), dashed (red) and dot-dashed (blue) lines, respectively.}
\label{fig:7}
\end{figure}
Here, the symbols $\sn$, $\sd$, $\cn$, and $\dn$ represent Jacobi elliptic functions, and 
\be
\bar{x}=\frac{1}{\sqrt{1+\alpha^2}}K\left(\frac{1}{\sqrt{1+\alpha^2}}\right).
\ee
Furthermore, there is another  solution, representing an anti-compact solution. The energy density associated to \eqref{comp1a} is
\ben \label{ecomp}
\rho(x) =  \left\{
\begin{array}{ll}
0, \,\, |x|>\bar{x},\\
 \dfrac{\cn^4\left(x \sqrt{1+\alpha^2} ,\dfrac{1}{\sqrt{1+\alpha^2}}\right)}{\dn^4\left(x \sqrt{1+\alpha^2},\dfrac{1}{\sqrt{1+\alpha^2}}\right)}, \,\, |x| \leq \bar{x}.
\end{array} \right.
\een
$U(z)$ inside the compact region $-\bar{x}/\sqrt{3}\leq z \leq \bar{x}/\sqrt{3}$ is
\ben \label{epc}
\dfrac{U(z)}{6\alpha^2 } &=&  \frac{\sn^2\left( z \sqrt{3(1+\alpha^2)},  \dfrac{1}{\sqrt{1+\alpha^2}}\right)\left[\sn^2\left(  z \sqrt{3(1+\alpha^2)}, \dfrac{1}{\sqrt{1+\alpha^2}}\right)+2\alpha^2\right] -(1+\alpha^2)}{(1+\alpha^2) \, \cn^2\left( z\sqrt{3(1+\alpha^2)},\dfrac{1}{\sqrt{1+\alpha^2}}\right)\dn^2\left(z\sqrt{3(1+\alpha^2)},\dfrac{1}{\sqrt{1+\alpha^2}}\right)}. \nonumber \\
\een
Outside the region in which the compact solution is localized, $U(z)$ is infinity and there is no propagating fluctuations.

In Fig. \ref{fig:7}, we illustrate the potential \eqref{vcomp1}, the compact solution \eqref{comp1a}, its energy density \eqref{ecomp} and the stability potential \eqref{epc}, for some values of $\alpha$.  Note that the  stability potential behaves as an infinite well supporting only bound states; and at the origin also occurs a change from a maximum to a minimum at $\alpha=\sqrt{2}$, as seen before in Fig.~\ref{fig:3} for the model with standard kinematics, pointing out that the modified dynamics does not alter this attribute. Moreover such behavior near the origin is not enough to split the energy density.  

\subsection{Second model}

Also, we can apply the deformation method extended to modified dynamic, using the procedure introduced in \cite{DMGD}. For that, we consider a new Lagrange density
\be\label{Lg2}
{\mathcal L}=-\frac{1}{4}\left(\partial_\mu \chi \partial^\mu \chi\right)^2-U(\chi).
\ee 
By means of the deformation function $\phi=g(\chi)$, the potentials of the systems described by  the Lagrange densities \eqref{Lg1}  and \eqref{Lg2}  are related by
\be
U(\chi)=\frac{V(\phi\rightarrow g(\chi))}{(dg/d\chi)^4}\,,
\ee
and the static solutions for the new model are given by the inverse function, 
$\chi(x)=g^{-1}(\phi(x))$.

We consider the deformation function 
\be
g(\chi)= \arcsinh\left(\sqrt{s}\,\tanh(\chi)\right)\,,
\ee
where  $s$ is real,  which 
applied to the potential \eqref{vcomp1} furnishes the new potential  
\ben\label{vcomp2}
 U(\chi)&=&\frac3{4s^2}\left(1+(s+1)\sinh^2(\chi)\right)^2 \left(1-(s\alpha^2-1)\sinh^2(\chi)\right)^2\,.
\een 
It requires that $s>1/\alpha^2$ and its static solutions are also compact-like
\be\label{comp2}
\chi(x)=\arctanh\left(\frac1{\sqrt{s}}\sinh(\phi(x))\right)\,,
\ee
where $\phi(x) $  is given by \eqref{comp1a}. Or yet,
\begin{eqnarray}\label{comp22}  
\chi(x) =   \left\{
\begin{array}{ll}
-\arctanh\left(\dfrac{1}{\alpha\sqrt{s}}\right), \, x<-\bar{x},\\ 
 \arctanh\left[\dfrac{1}{\sqrt{s(1+\alpha^2)}} \sd\left( x \sqrt{1+\alpha^2},\dfrac{1}{\sqrt{1+\alpha^2}}\right)\right], \, |x|\leq\bar{x},\\
\arctanh\left(\dfrac{1}{\alpha\sqrt{s}}\right), \,x>\bar{x}\,.
\end{array} \right. 
\end{eqnarray}
The energy density is
\ben 
\rho(x) =  \left\{
\begin{array}{ll}
0,\:\:\:\:\:\: |x|>\bar{x},\\
\dfrac{s^2\, (1+\alpha^2)^4 \, \cn^4\left(x \sqrt{1+\alpha^2},\frac{1}{\sqrt{1+\alpha^2}}\right)}{\dn^8\left(x \sqrt{1+\alpha^2},\frac{1}{\sqrt{1+\alpha^2}}\right)\left[s (1+\alpha^2)-\sd^2\left(x \sqrt{1+\alpha^2},\frac{1}{\sqrt{1+\alpha^2}}\right)\right]^4}, \:\:\:\:\:\: |x| \leq \bar{x}.
\end{array} \right. \nonumber \\
\label{dec2}
\een

\begin{figure}
\centering
\includegraphics[width=7cm,height=6.5cm]{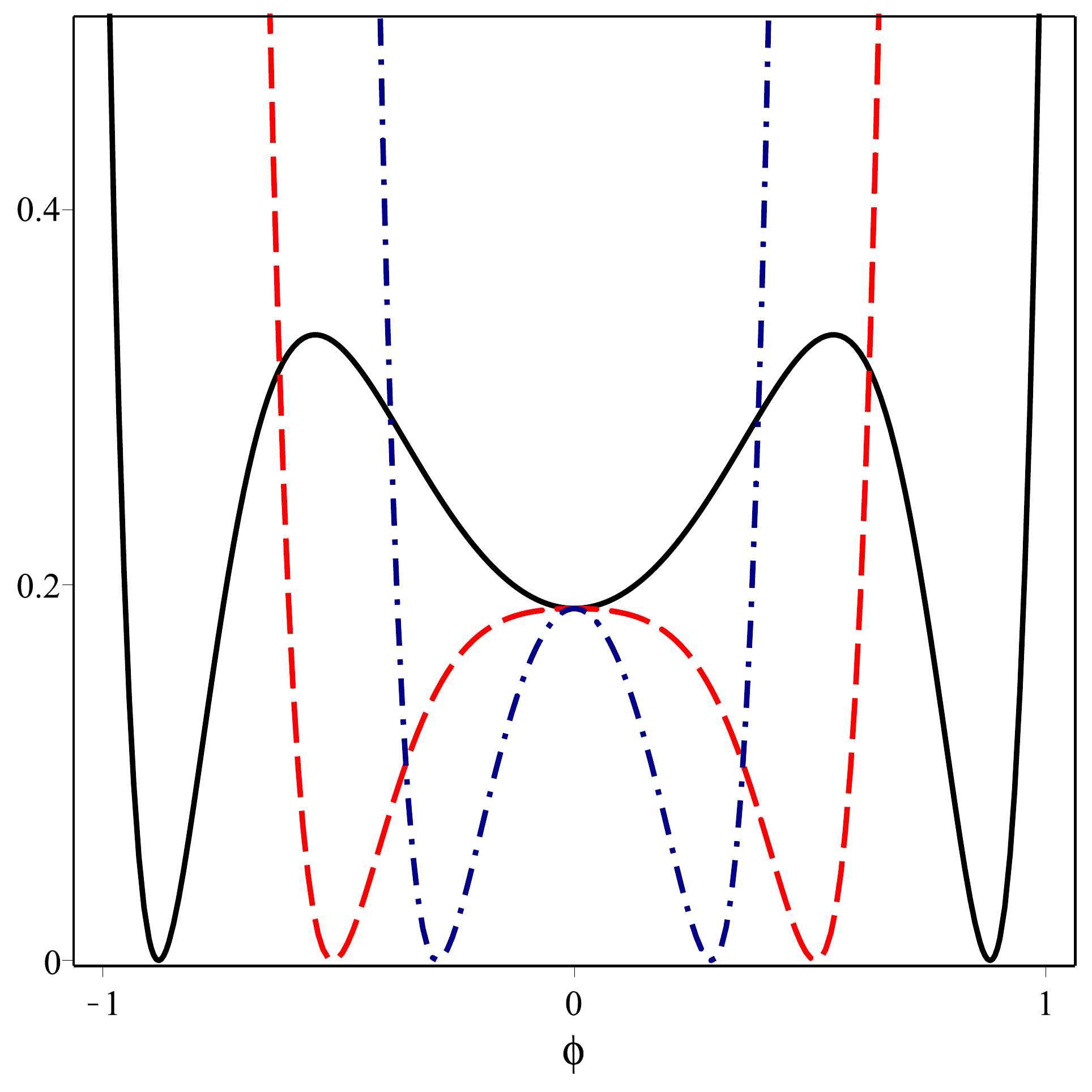}
\includegraphics[width=7cm,height=6.5cm]{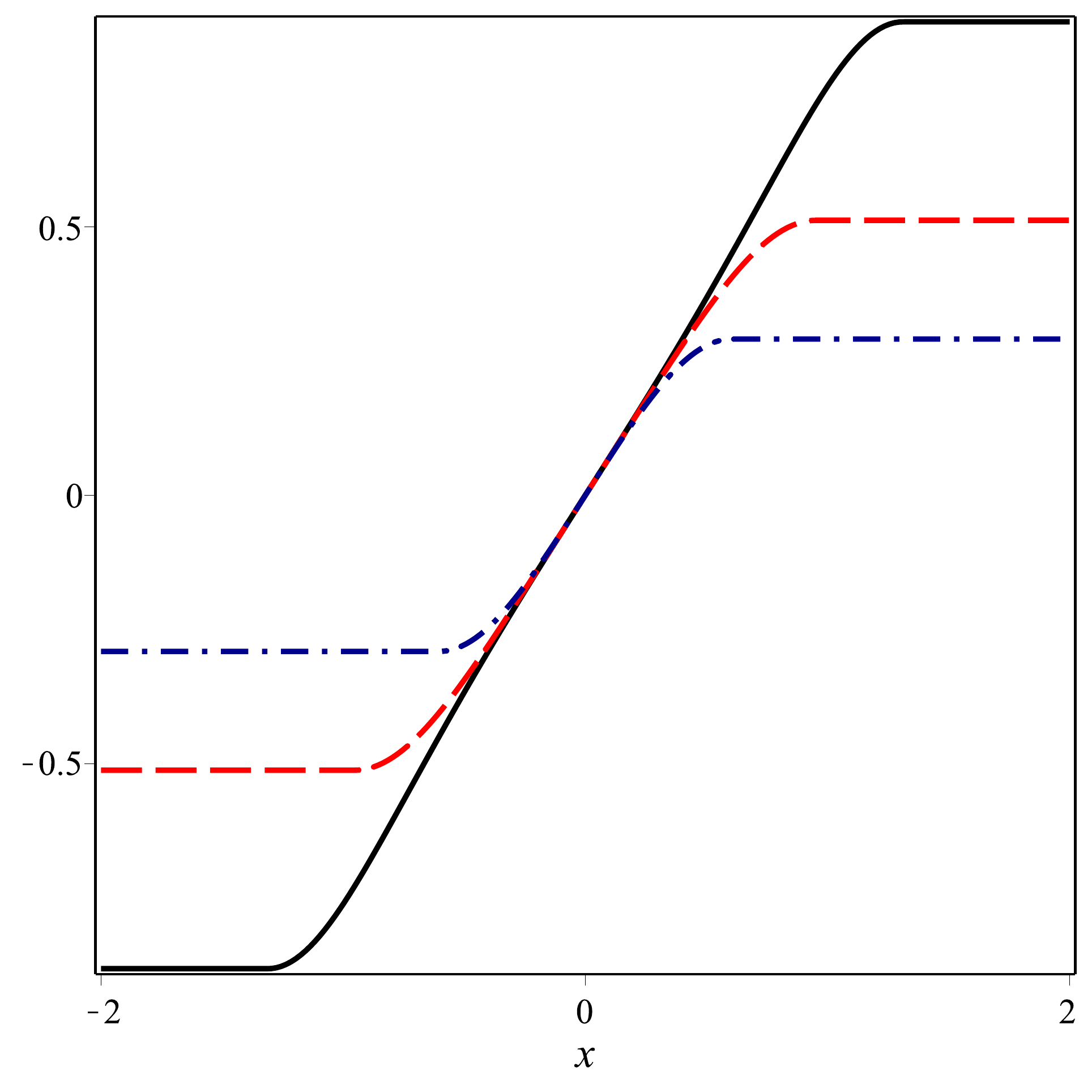}
\includegraphics[width=7cm,height=6.5cm]{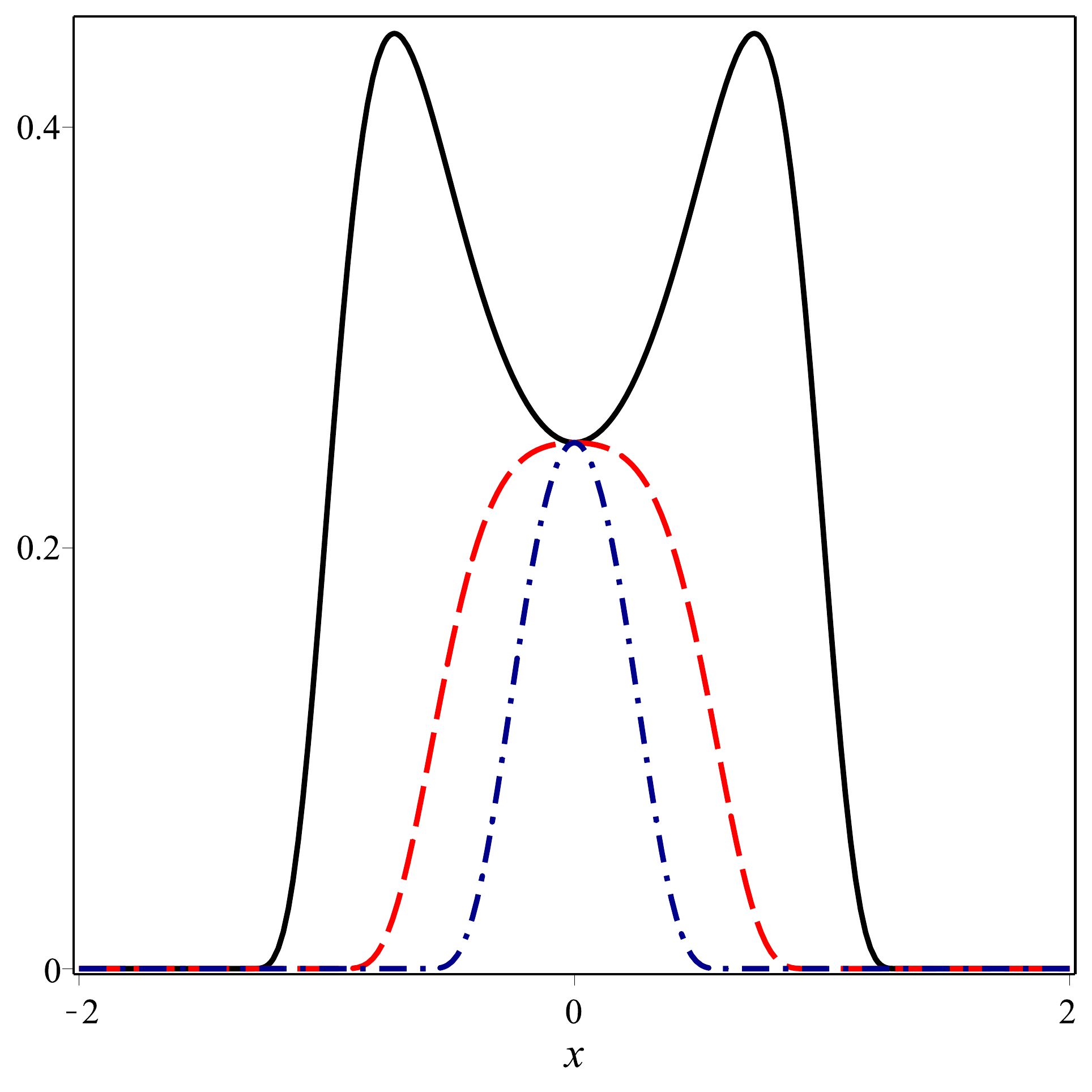}
\includegraphics[width=7cm,height=6.5cm]{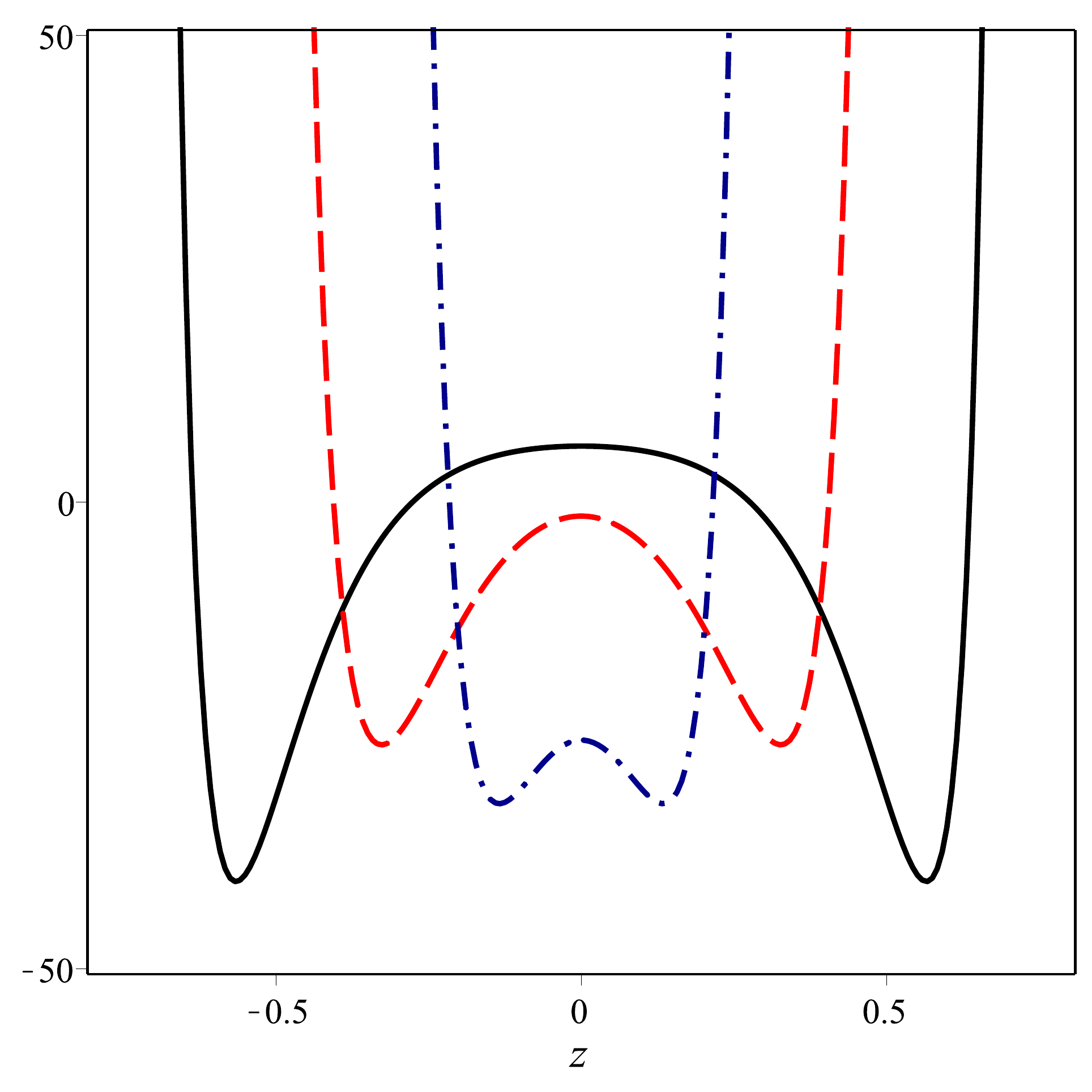}
\caption{In the top panel one depicts the potential \eqref{vcomp2} (left) and the compact solution \eqref{comp22} (right). In the bottom panel one shows the energy density \eqref{dec2} (left) and the stability potential \eqref{epof} (right) obtained for the compact kink \eqref{comp22}. We are using ${s=2}$ and varying ${\alpha=1, 3/2, 5/2}$, depicting the curves with solid (black), dashed (red) and dot-dashed (blue) lines, respectively.}
\label{fig:8}
\end{figure}

In Fig. \ref{fig:8}, we illustrate the potential \eqref{vcomp2},  as well as the compact kink \eqref{comp22}, the energy density \eqref{dec2}  and the stability potential which comes through \eqref{epof} and produces an awkward expression which we omit in the current work. Here, the energy density is characterized by a splitting of the maximum into two new maxima, which vanishes at $\alpha^2=(s+2)/2$, as shown in Fig.~\ref{fig:8}. See also that the stability potential $U(z)$ has a plateau near the origin, which disappears at $\alpha^2=4+6/s+\sqrt{15+36/s+22/s^2}$. 

\begin{figure}
\includegraphics[width=7cm,height=6.5cm]{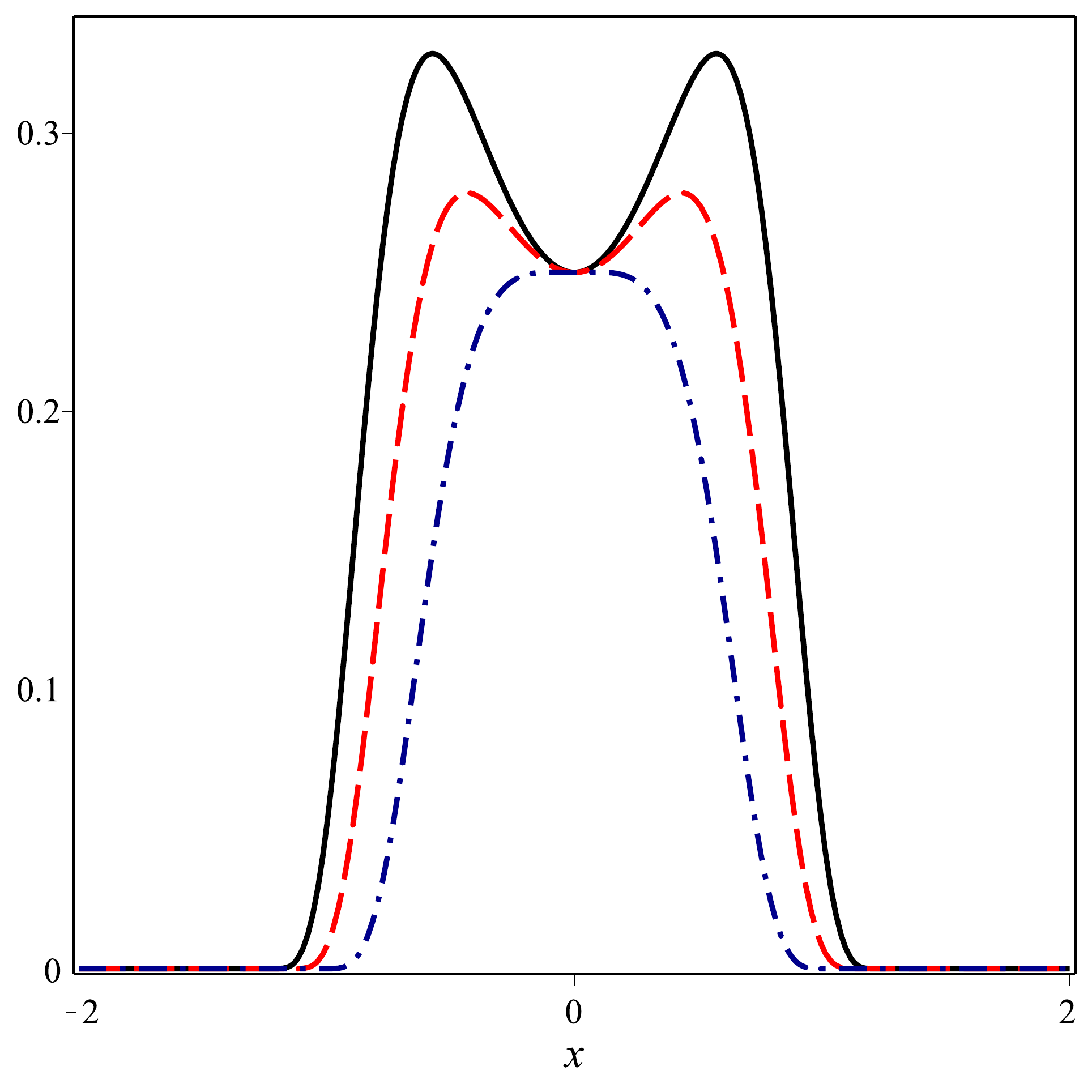}
\includegraphics[width=7cm,height=6.5cm]{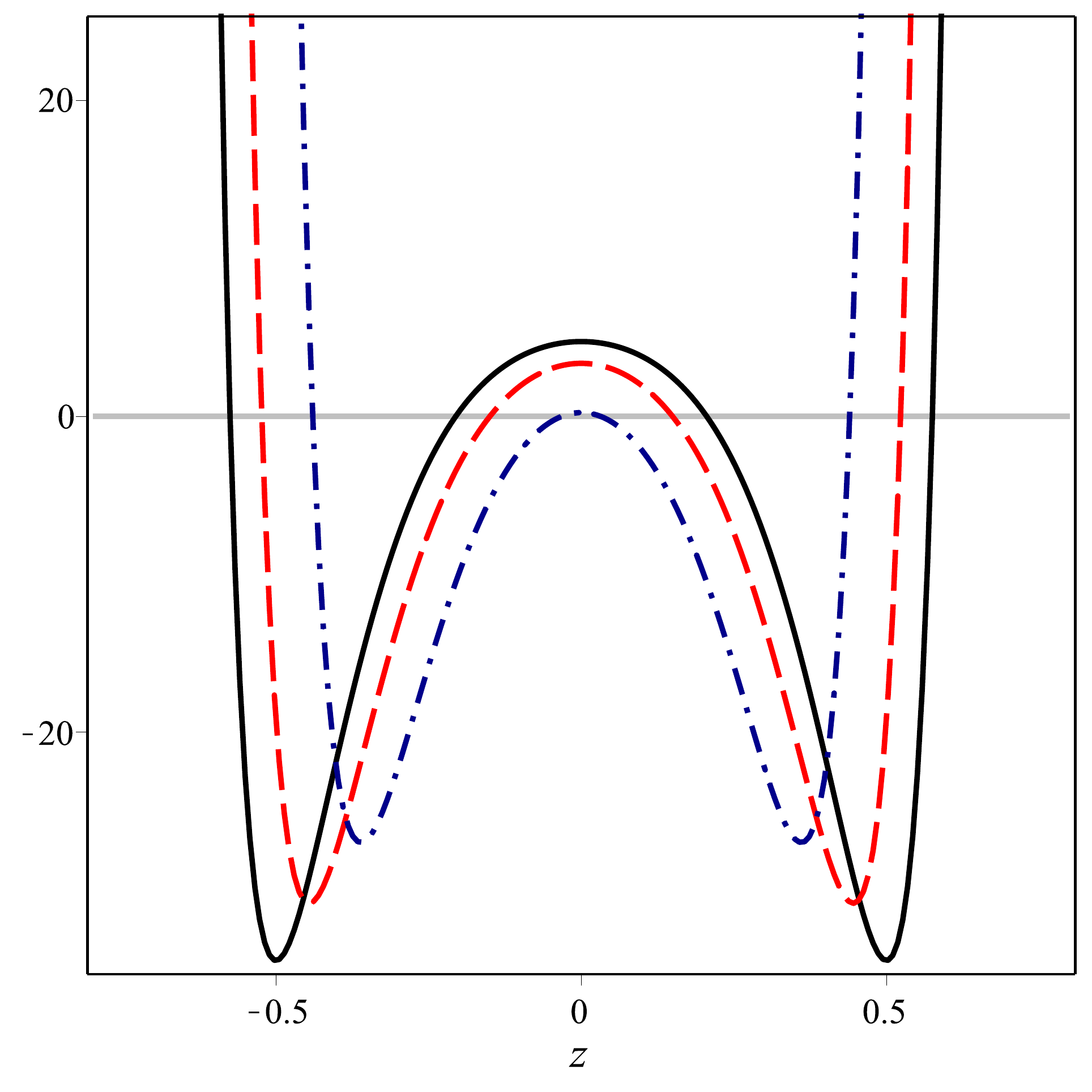}
\caption{The energy density \eqref{dec2} (left panel) and  the stability potential \eqref{epof} calculated for the compact kink \eqref{comp22} (right panel). They are depicted for $s=2$ and for $\alpha=1.1, 1.2, 1.4$, with solid (black), dashed (red) and dot-dashed (blue) lines, respectively.}
\label{fig:9}
\end{figure}

In order to further explore the splitting phenomenon, we display in Fig.~\ref{fig:9} the energy density \eqref{dec2} and  the corresponding stability potential \eqref{epof} obtained for the compact kink \eqref{comp22}. Here we take values of the parameter $\alpha$  which highlight the splitting phenomenon. Differently from the previous case illustrated in Fig.~\ref{fig:7}, the plateau on the stability potential is capable of inducing the  splitting on the energy density. This behavior also appeared on braneworld scenarios with generalized gravity \cite{Lobao}, or yet in scenarios with standard gravity where thermal effects \cite{Campos} or the presence of internal structure \cite{2kink1,2kink2} are responsible by the splitting. Another system which induces the splitting is found in the model studied in Ref.~\cite{livre}.

Since the splitting depends on the values of the parameters that specify the models, we believe that this should be better investigated in connection with the issue which concerns the configurational entropy of the structure, to see how it changes as we vary the parameters of the model. This possibility can be implemented following the lines of the recent works \cite{CE1,CE2,CE3} and we hope to report on this in the near future.

\section{Braneworld models}\label{sec-4}

Applications can be found from a modification in the geometry of the system. We incorporate the scalar field into a five-dimensional warped geometry with a single extra dimension of infinite extent described by the line element \cite{RS,Fre}
\be
ds_5^2=\e^{2A}\eta_{\mu\nu}dx^{\mu}dx^{\nu}-dy^2,
\ee
where $A = A(y)$ is the warp factor, $y$ is the extra dimension, and $\eta_{\mu\nu}$ describes the  four-dimensional Minkowski spacetime $(\mu, \nu = 0, 1, 2, 3)$. In this case, the action is written as
\be
\label{action}
I=\int{d^4xdy\sqrt{|g|}}\left(-\frac{R}{4}+{\cal L}(\phi,\partial_{\mu}\phi)\right).
\ee
where $R$ is the scalar curvature and ${\cal L}(\phi,\partial_{\mu}\phi)$ describes the scalar field.

\subsection{Standard model} 

In the standard formalism, the Lagrange density ${\cal L}(\phi,\partial_{\mu}\phi)$ is defined by \eqref{lagran}. We assume that the scalar field only depends on the extra dimension $y$. For these reasons, the equation of motion for $\phi$ and the Einstein's equations are expressed as
\bens
\phi''+4A'\phi'&=&V_{\phi}, \\
A''&=&-\frac{2}{3}\phi'^2, \\
A'^2&=&\frac{1}{6}\phi'^2-\frac{1}{3}V.
\eens
It is useful work with a first order formalism, such that if we write the potential as
\be
\label{pot5}
V(\phi)=\frac{1}{2}W_{\phi}^2-\frac{4}{3}W^2.
\ee
then the equations of motion are reduced to 
\ben
\label{qq}
\frac{d\phi}{dy} &=& W_{\phi}, \\
\label{qq2}
\frac{dA}{dy} &=& -\frac{2}{3}W(\phi).
\een
In this case, the energy density has the form
\be
\label{eneb}
\rho(y)=\e^{2A}\left(W_{\phi}^2-\frac{4}{3}W^2\right).
\ee

\begin{figure}
\includegraphics[width=7.5cm,height=6cm]{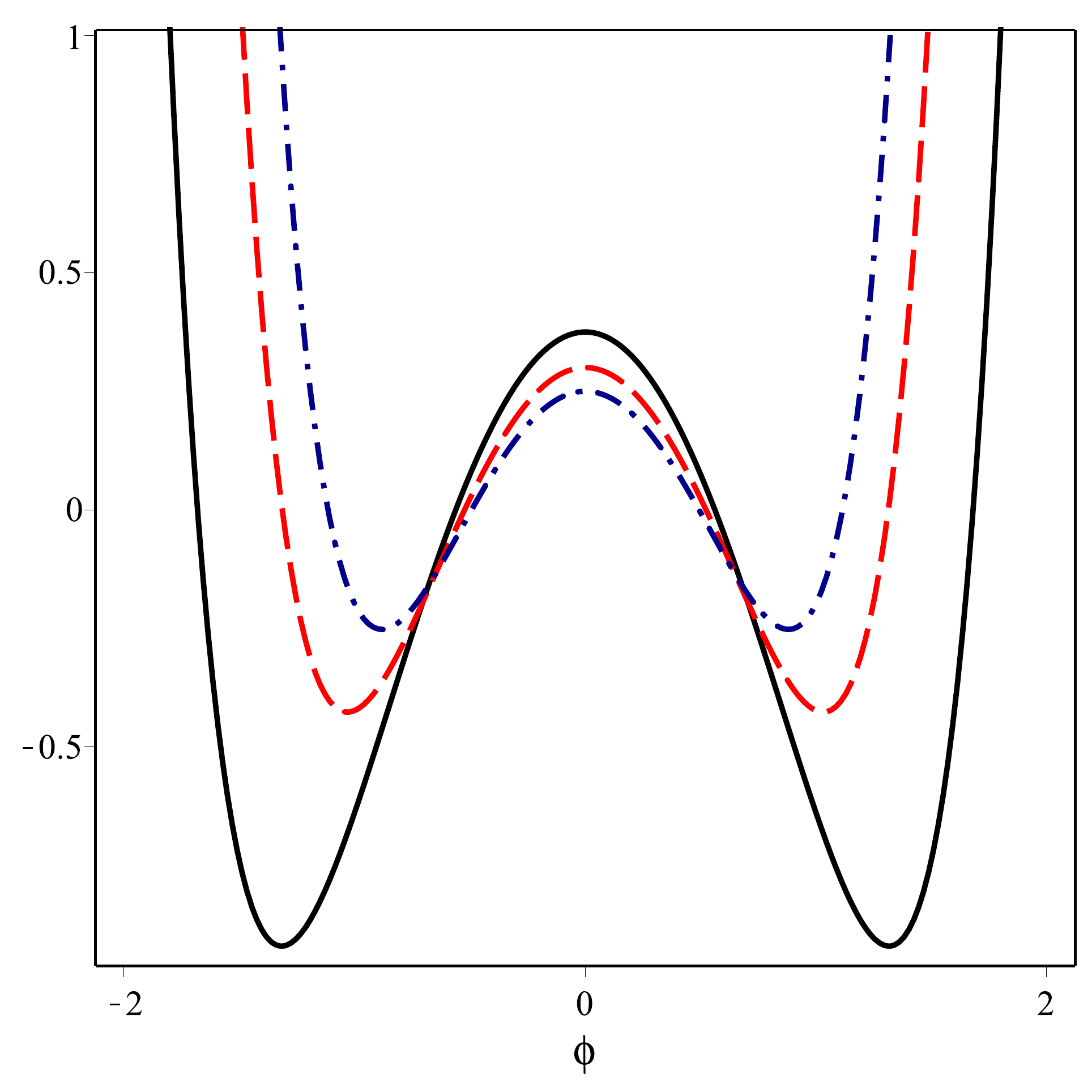}
\includegraphics[width=7.5cm,height=6cm]{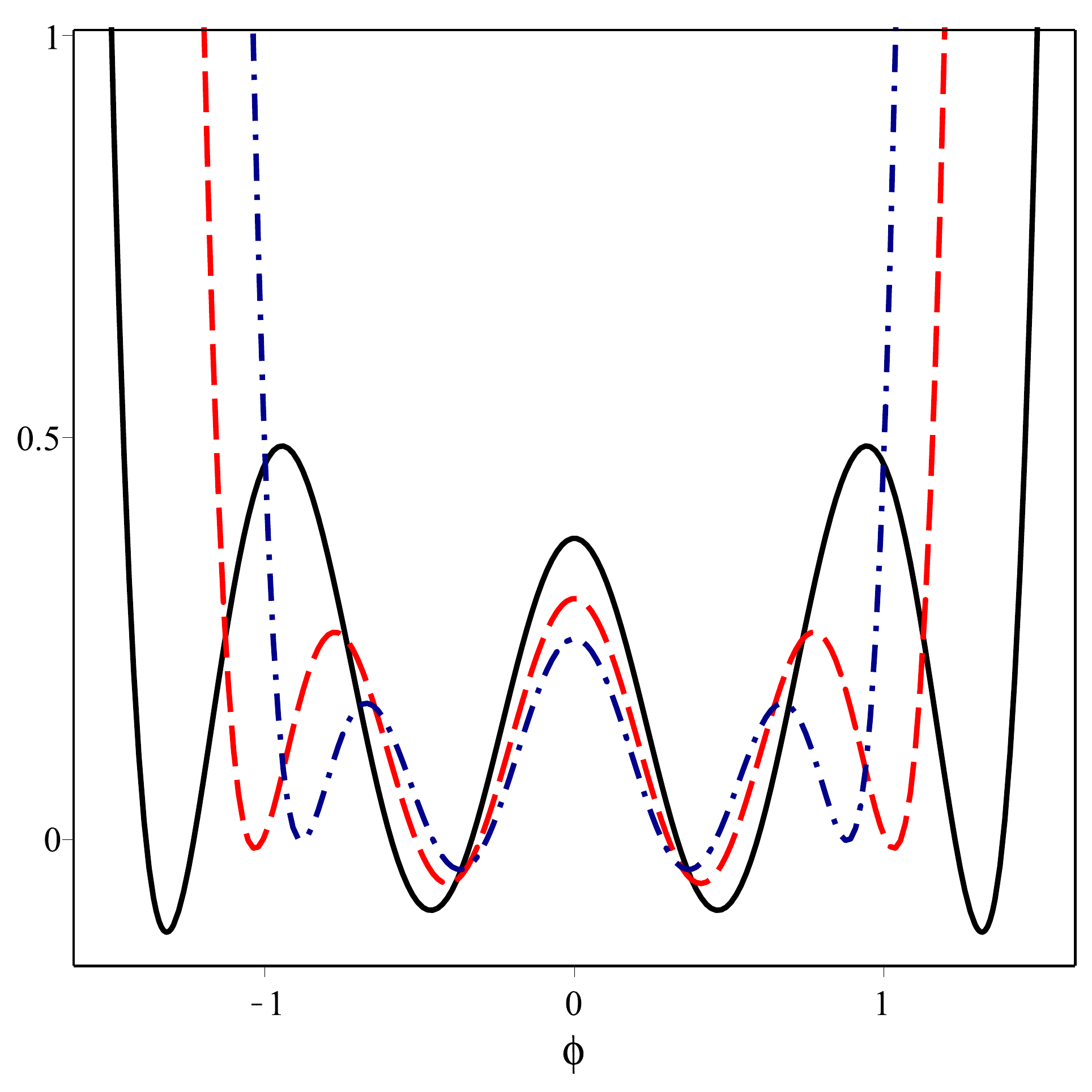}
\caption{The potential \eqref{pot5} is depicted for $n=4/3,5/3,2$, with solid (black), dashed (blue) and dot-dashed (red) lines, respectively.
In the left panel one takes $a=1$ and in the right panel, $a=3$.}
\label{fig:10}
\end{figure}

We now turn attention to the family of potentials with $a$-odd, developed in \ref{family-1}, in order to introduce braneworld models. So we use $W$ in the form
\ben
W_a^{(1)}&=&\frac{1}{\sqrt{n}a}\int d\phi \,U_{a-1}(\sqrt{n}\tanh(\phi)) 
\left[1-(n-1)\sinh^2(\phi)\right].
\een
Explicitly, for $a=1,3$ we have 
\bens
W_1^{(1)}&=&\frac{(1+n)}{2\sqrt{n}}\phi-\frac{(n-1)}{2\sqrt{n}}\cosh(\phi)\sinh(\phi), \\
W_3^{(1)}&=&\frac{(12n^2-5n-1)}{6\sqrt{n}}\phi-\frac{(n-1)(4n-1)}{6\sqrt{n}} \cosh(\phi)\sinh(\phi)-\frac{4}{3}n^{3/2}\tanh(\phi). \:\:\:\:\:\:\:\:\:\:\:\:
\eens
The potentials defined by these models, $W_1^{(1)}$ and  $W_3^{(1)}$, are plotted in Fig.~\ref{fig:10} assuming different values of $n$.

\begin{figure}
\includegraphics[width=7.5cm,height=6.5cm]{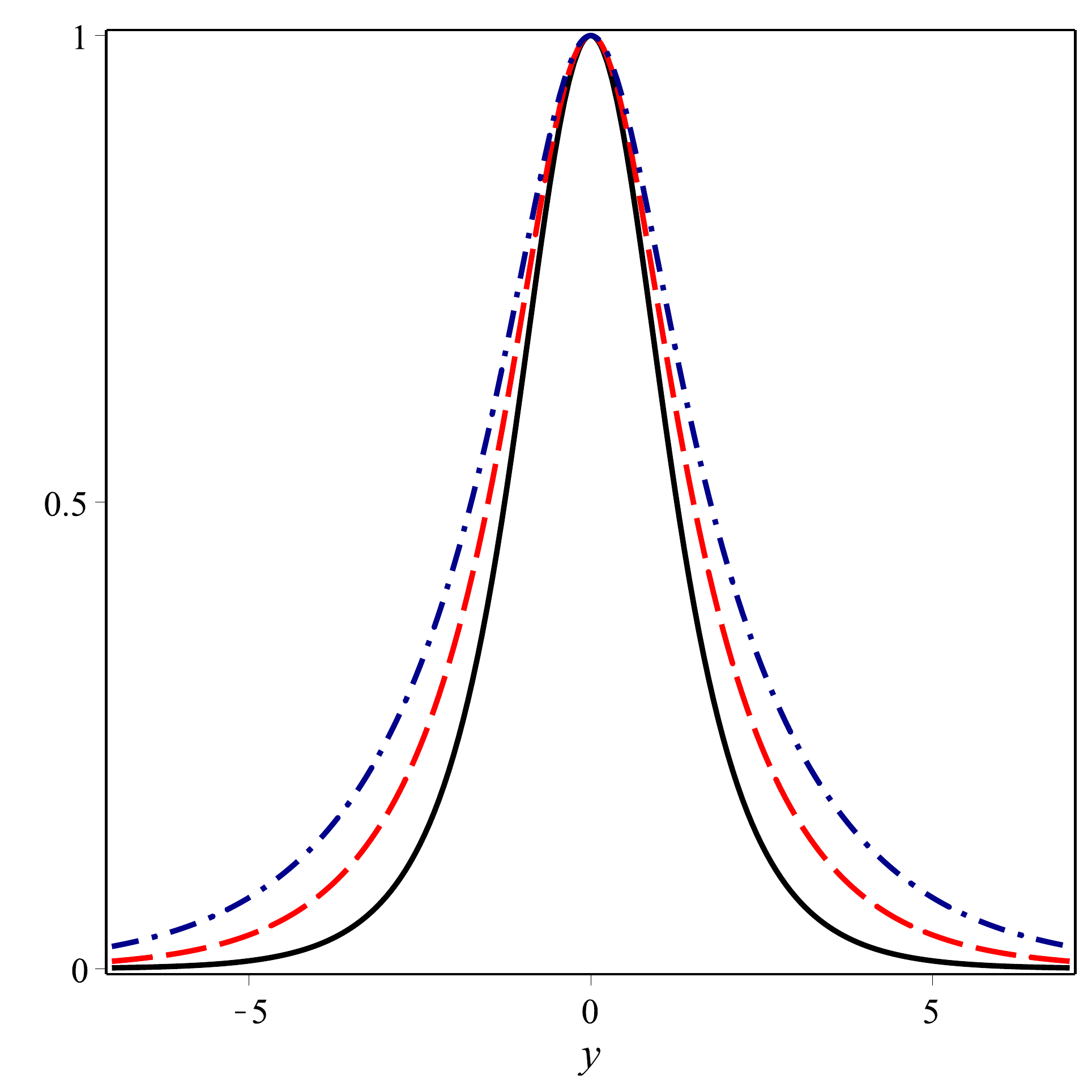}
\includegraphics[width=7.5cm,height=6.5cm]{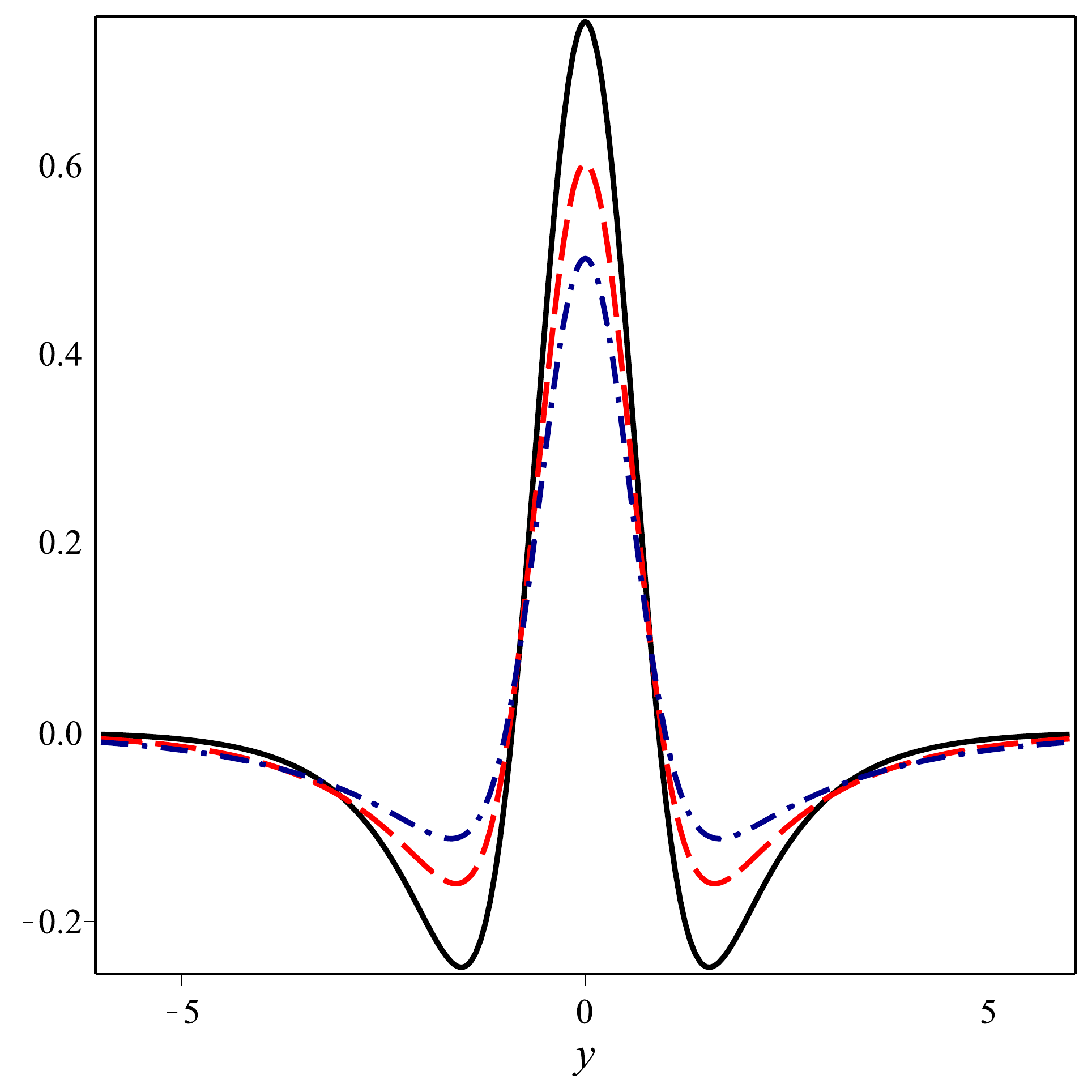}
\caption{The warp factor $\e^{2A}$ (left) and energy density $\rho$ (right) for the model defined by $W_1^{(1)}$, depicted for $n$ as in
Fig.~\ref{fig:10}.}
\label{fig:11}
\end{figure}

\begin{figure}
\includegraphics[width=7.5cm,height=6.5cm]{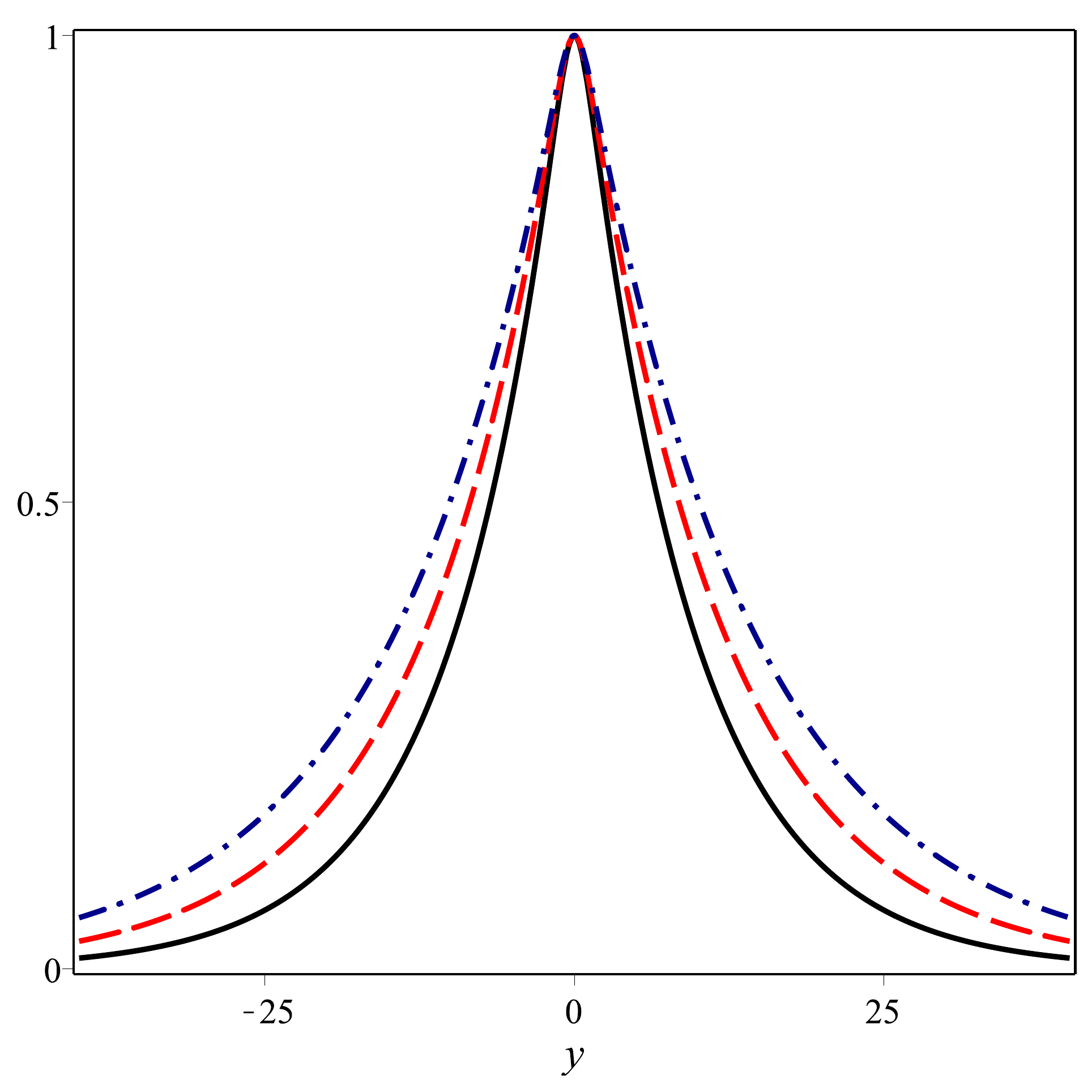}
\includegraphics[width=7.5cm,height=6.5cm]{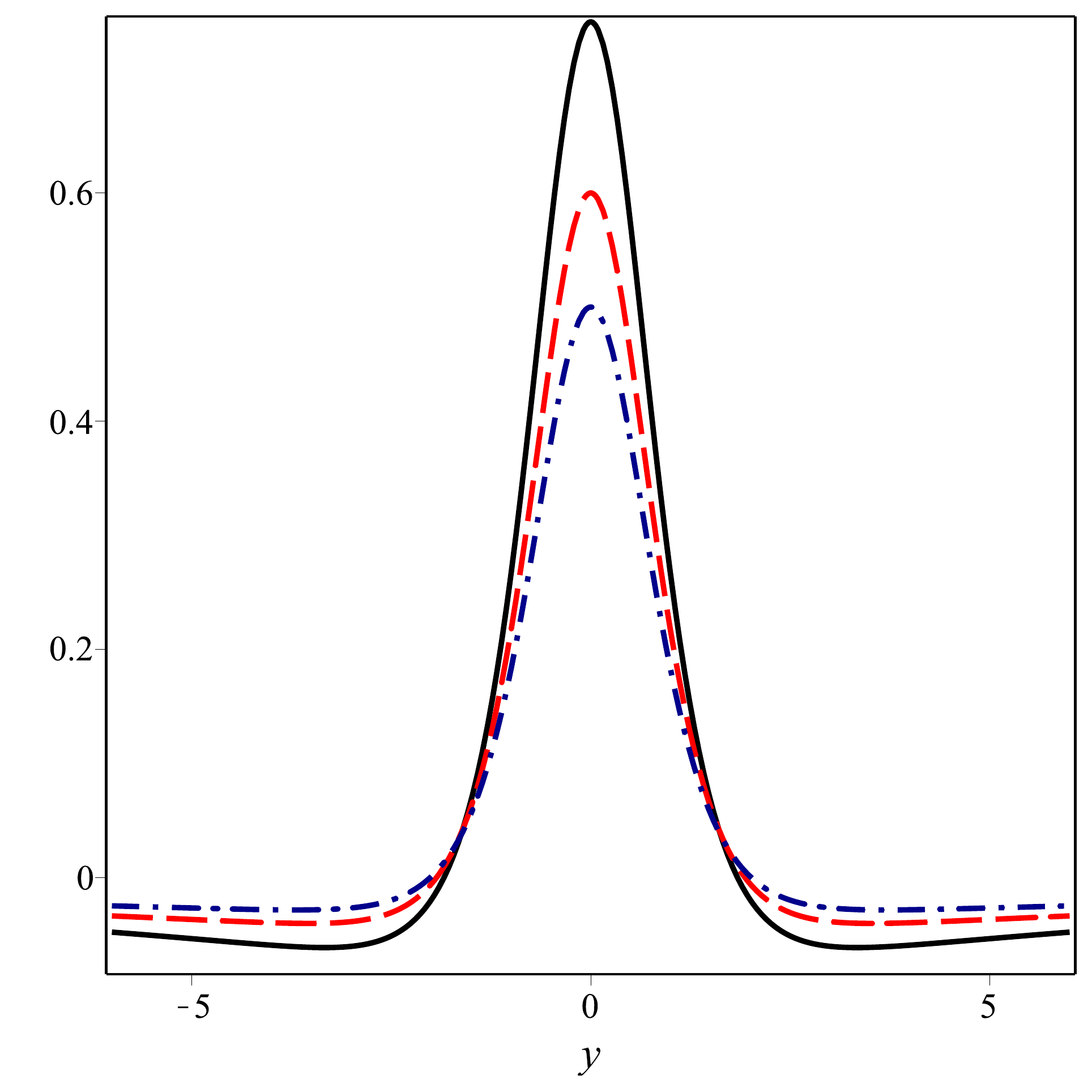}
\caption{The warp factor $\e^{2A}$  (left) and energy density $\rho$ (right) for the model defined by $W_3^{(1)}$, with the same values of $n$ presented in Fig.~\ref{fig:10}. The results represent the central sector of the potential the appears in the right panel of Fig.~\ref{fig:10}.}
\label{fig:12}
\end{figure}

The scalar field solutions are similar to the ones previously obtained in Eq.~\eqref{sol}, for the flat spacetime.  Moreover, we solve numerically the first-order equation \eqref{qq2} to get the warp factor and we calculate the energy density \eqref{eneb}, for $a=1,3$. These results are presented in Figs.~\ref{fig:11} and \ref{fig:12}. We only consider the cases of symmetric brane, but in the case of $a=3$, the potentials may generate one symmetric brane and two asymmetric branes (one related to the left sector and the other to the right sector of $V$). The asymmetric warp factors can become divergent depending on the value of $n$. Similar results are found for others values of $a$, in the odd case. The cases of asymmetric brane are similar to the studies implemented in \cite{Aw1,Aw2} and in references therein. We will not discuss this possibility in the current work.


\subsection{Non-standard model}

Let us now focus on a flat brane driven by a scalar field with the non-standard kinetic term, where ${\cal L}(\phi,\partial_{\mu}\phi)$ is defined by  \eqref{Lg1}. For  $\phi = \phi(y)$,  the equation of motion for the scalar field and the Einstein's equations are
\bens
3\phi'^2\phi''+4A'\phi'^3&=&V_{\phi}, \\
A''&=&-\frac{2}{3}\phi'^4, \\
A'^2&=&\frac{1}{4}\phi'^4-\frac{1}{3}V.
\eens
For simplicity, we adopt the first order formalism developed in Ref.~\cite{BGD}, where the equations for $\phi'$ and $A'$ become
\be
\label{bforq}
\phi'=W_{\phi}^{1/3}, \:\:\:\: A'=-\frac23 W(\phi). 
\ee
And the potential has the form
\be
\label{pbgd}
V(\phi)=\frac{3}{4}W_{\phi}^{4/3}-\frac{4}{3}W^2.
\ee
Moreover, the energy density is
\be
\label{enebb}
\rho(y)=\e^{2A}\left(W_{\phi}^{4/3}-\frac{4}{3}W^2\right).
\ee

\begin{figure}
\centering
\includegraphics[width=7cm,height=6.5cm]{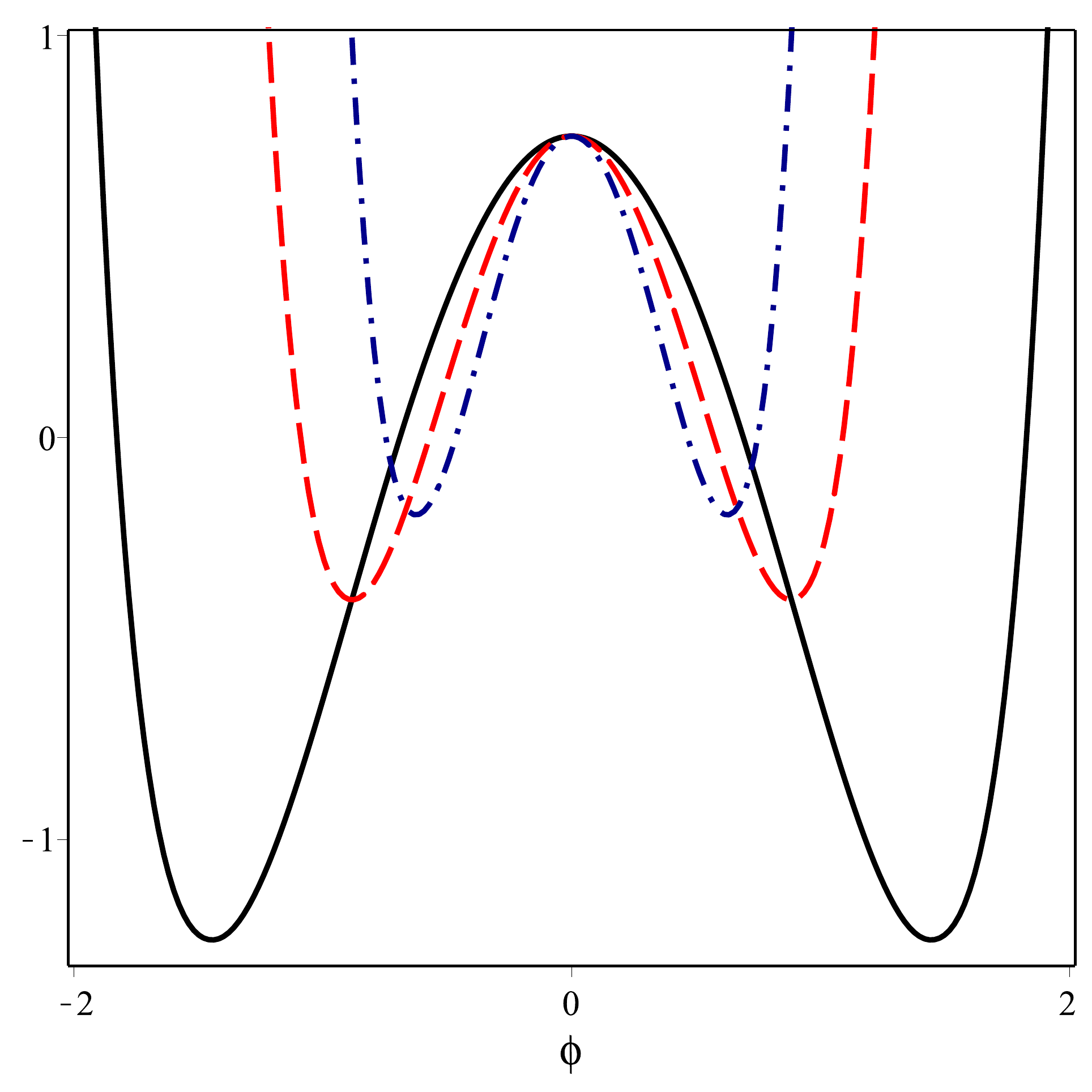}
\includegraphics[width=7cm,height=6.5cm]{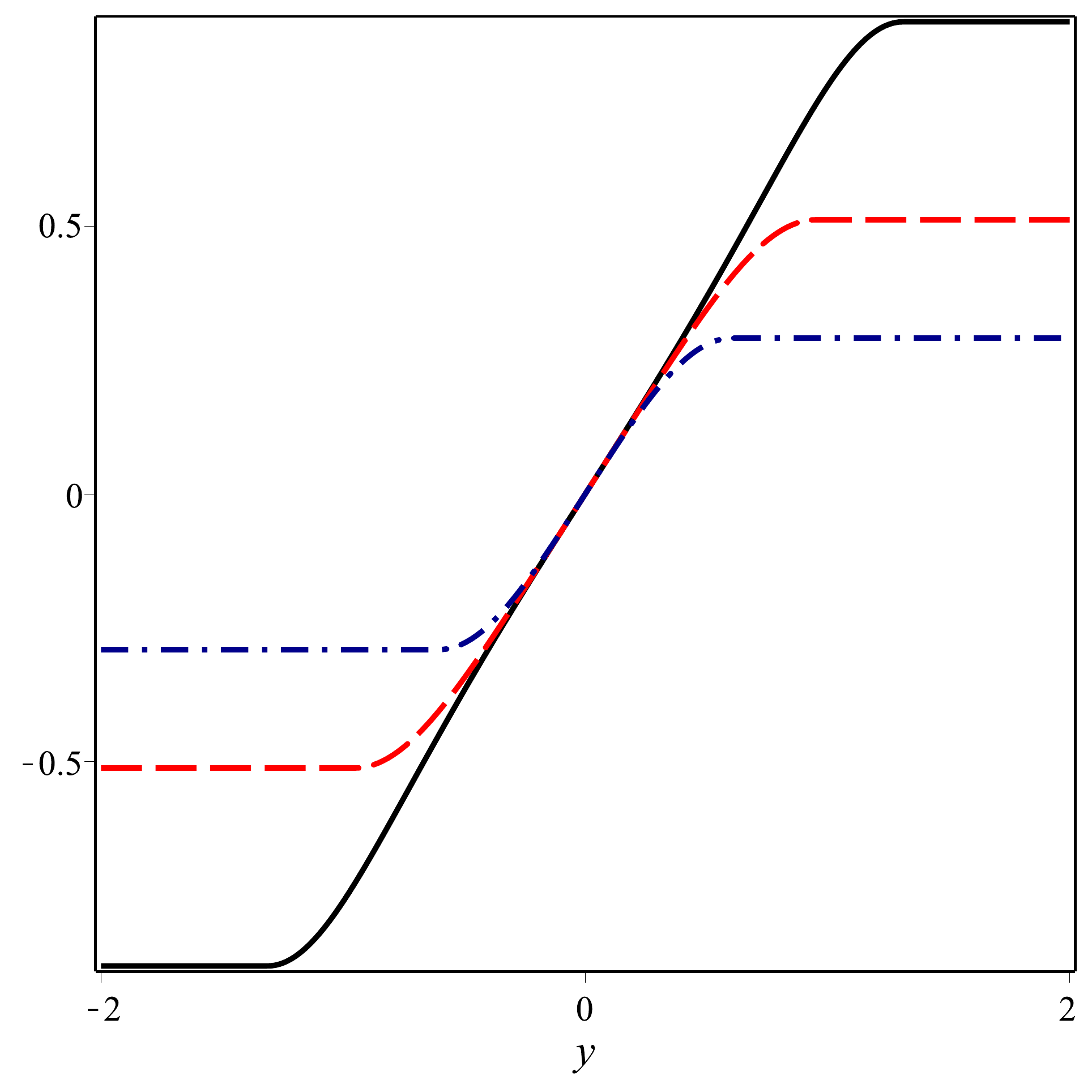}
\includegraphics[width=7cm,height=6.5cm]{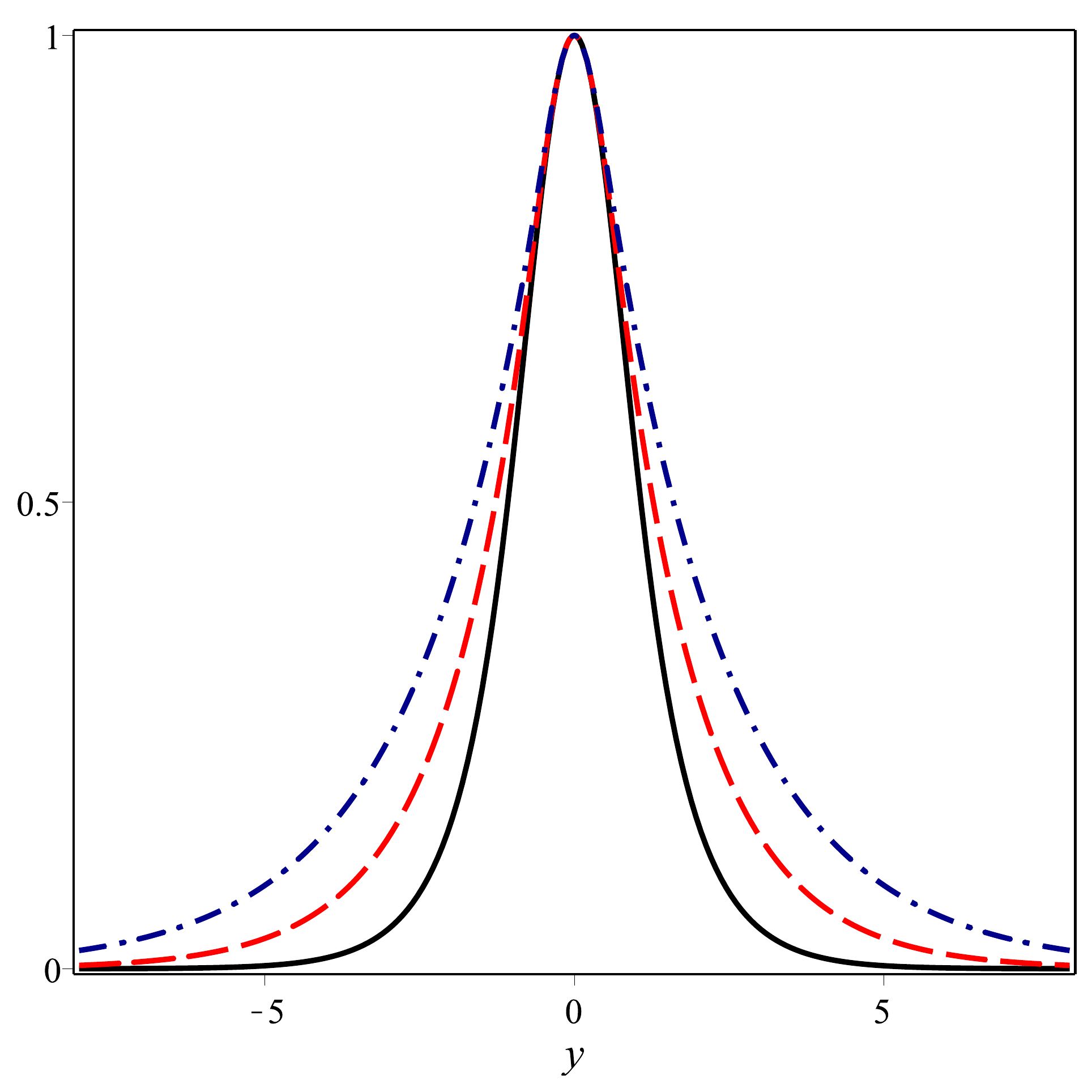}
\includegraphics[width=7cm,height=6.5cm]{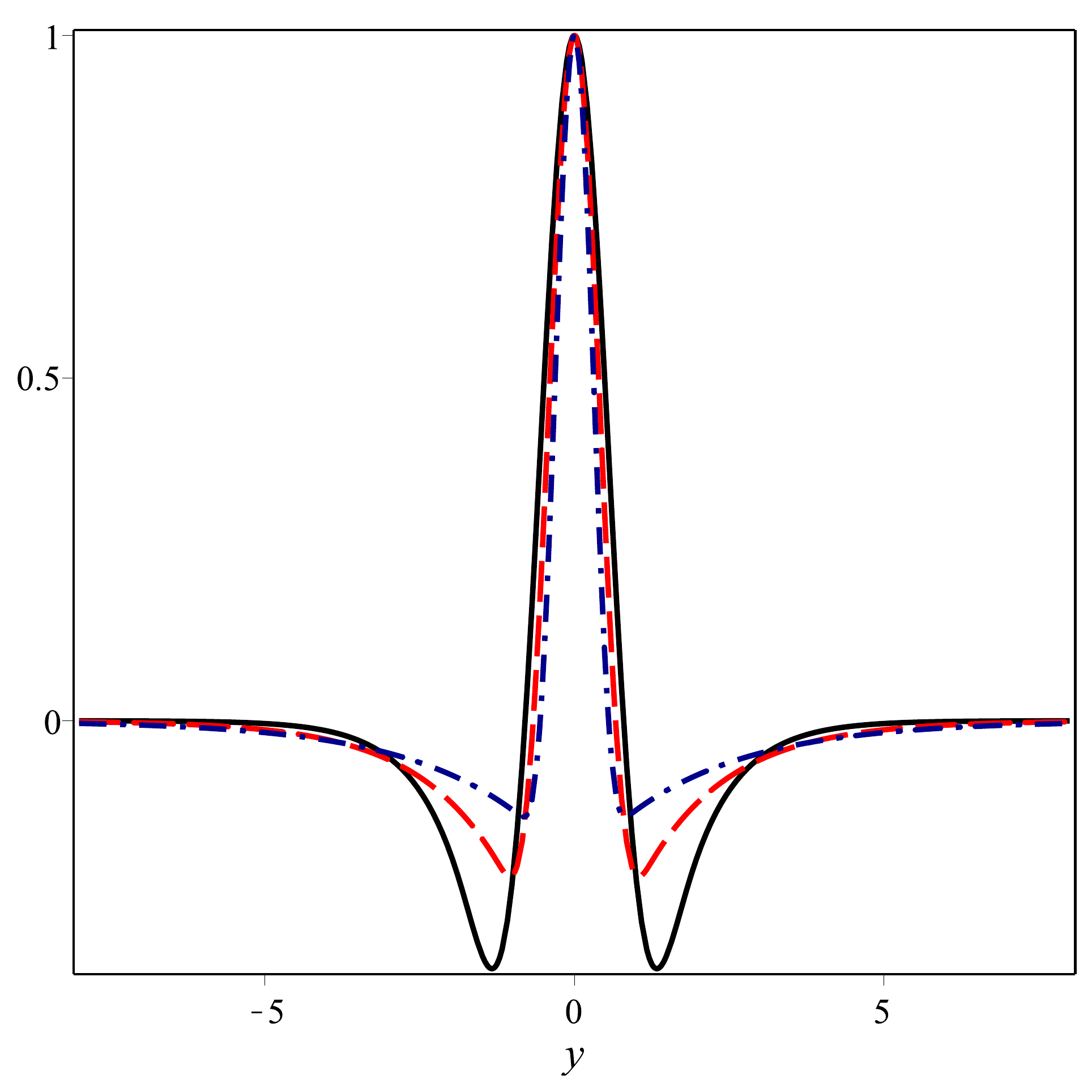}
\caption{In the top panel one shows the potential \eqref{pbgd} (left) and the compact solution \eqref{comp1a} (right). In the bottom panel one displays the numerical solution for the warp factor \eqref{bforq} (left) and the energy density \eqref{enebb} (right). We are using $\alpha=1/2,1,3/2$, represented by solid (black), dashed (red) and dot-dashed (blue) lines, respectively.}\label{fig:13}
\end{figure}

We consider the case 
\be
W_{\phi}=\left(1-\alpha^2\sinh^2(\phi)\right)^{3/2},
\ee
and now the solution for the scalar field remains as in Eq.~\eqref{comp1a} and the warp factor is obtained numerically by the first-order equation \eqref{bforq}. In Fig. \ref{fig:13}, we display the potential, the kink solution, the warp factor and the energy density.

As we have just seem, the first order formalism can be implemented for both the standard and non-standard models. The presence of the first order formalism allows to inform that the two scenarios are stable against tensorial fluctuations in the gravity sector, so the standard and non-standard models that we have just investigated are robust against fluctuation of the metric tensor \cite{Fre,BGD}.

\section{Comments and conclusions}
\label{sec-com}

In this work we constructed two distinct families of scalar field models, described by hyperbolic interactions in the case of standard kinematics.  The static solutions for these models were obtained from the deformation procedure developed in Ref.~\cite{AA1,AA2,AA3}, and we have studied their energy densities, stability potentials and the corresponding zero modes. 

These families of models are governed by hyperbolic interactions. They are new models that require further investigations. A particularly interesting issue concerns the behavior of the their kinklike configurations under collisions, in a way similar to the recent studies already implemented in Refs.~\cite{C0a,C0b,C1a,C1b,C2,C3,C4}. These works study the kink-antikink and simultaneous multi-kink collisions in several distinct models, including polynomial and non polynomial interactions, but the models that we include and solve in the current work are all new and may inspire new investigations on how their solutions behave under collisions. Since the sinh-Gordon model is directly connected to minimal surfaces \cite{MS1a,MS1b,MS2} and integrability \cite{IN1,IN2,IN3}, the models that we investigate in this work may contribute to generate new effects in their kink-antikink and multi-kink collisions. 

We have also implemented a generalization, changing the kinematic term to a non-canonical term, and there we have seen that such hyperbolic models can support compact solutions. We have studied the stability potential, identifying some peculiar characteristics such as the appearance of a local maximum at the origin, which depends on the value of the real parameter there introduced. To go on, we have used the deformation procedure developed for the case of non-standard dynamics \cite{DMGD}, and this enabled us to find another model with compact solutions, driven by a new hyperbolic potential.

It is important to notice that in the two cases, considered in Sec.~\ref{sec-2} and in Sec.~\ref{sec-3}, respectively, the use of the deformation procedure for standard kinematics \cite{AA1,AA2,AA3} and for non-canonical kinematics \cite{DMGD}, helped us to suggest and investigate the new models analytically. This fact motivate us to go further and investigate the splitting found in Fig.~\ref{fig:9}. Here, we think that a closer inspection on the configurational entropy may shed light on the value of $\alpha$ and on the presence of the splitting. 

Since the presence of kinks and compact kinks can be used to construct braneworld models in a five-dimensional warped geometry with a single extra dimension of infinite extent, we applied our results to the study of branes. The present investigation offers a systematic way to find defect solutions of new models described by a real scalar field with nontrivial interactions of the hyperbolic type. The results of the current work may inspire new researches on kinks and branes and, in particular, in the cases of minimal surfaces \cite{MS1a,MS1b,MS2}, integrability \cite{IN1,IN2,IN3}, and on the configuration entropy of the solutions \cite{CE1,CE2,CE3}.

Another issue of current interest is inspired in the AdS/CFT correspondence and concerns studies in the lines of \cite{HC1,HC2} and in references therein, and also connected to string cosmology, as recently reviewed in Ref.~\cite{B}. On the other hand, the case of generalized models can be extended to other possibilities, in particular to the case where the kinematics is changed to obey the Born-Infeld suggestion \cite{BI0a,BI0b}, with the square root constraining the field evolution and bringing new interesting effects. This can be implemented both in the flat spacetime, as it was for instance studied in \cite{BI}, or in curved spacetime, as in \cite{BI1,BI2}. We hope to report on these and other related issues in the near future.

\acknowledgments

This work is partially supported by CNPq, Brazil. DB acknowledges support from projects 455931/2014-3 and 306614/2014-6, EEML acknowledges support from project 160019/2013-3, and LL acknowledges support from projects 307111/2013-0 and 447643/2014-2.

\end{document}